\documentclass[aps,pra,preprint,amsmath,amssymb,floatfix,superscriptaddress,showpacs]{revtex4}

\usepackage{graphicx}
\usepackage{array}

\begin{document}

\title{Numerical Study of Amplified Spontaneous Emission\\ 
and Lasing in Random Media}

\author{Jonathan Andreasen}
\altaffiliation[Currently at]{
  Laboratoire de Physique de la Mati\`ere Condens\'ee, CNRS UMR 6622,
  Universit\'e de Nice-Sophia Antipolis,
  Parc Valrose, F-06108 Nice Cedex 02, France
}
\affiliation{
  Department of Applied Physics, Yale University,
  New Haven, Connecticut 06520, USA
}
\author{Hui Cao}
\affiliation{
  Department of Applied Physics, Yale University,
  New Haven, Connecticut 06520, USA
}
\affiliation{
  Department of Physics, Yale University,
  New Haven, Connecticut 06520, USA
}
\date{\today}

\begin{abstract}
  We simulate the transition from amplified spontaneous emission (ASE)  
  to lasing in random systems with varying degrees of mode overlap.
  This is accomplished by solving the stochastic Maxwell-Bloch equations with the finite-difference time-domain method. 
  Below lasing threshold, the continuous emission spectra are narrowed by frequency-dependent amplification.  
  Our simulations reproduce the stochastic emission spikes in the spectra.
  Well-defined peaks, corresponding to the system resonances, emerge at higher pumping
  and are narrowed by stimulated emission before lasing takes place.
  Noise tends to distribute pump energy over many modes, resulting in multi-mode operation.
  Well above the lasing threshold, the effects of noise lessen and results become
  similar to those without noise.
  By comparing systems of different scattering strength, we find that weaker scattering extends 
  the transition region from ASE to lasing, where the effects of noise are most significant.
\end{abstract}


\maketitle

\section{Introduction}

A random laser has two elements: an active material that spontaneously emits light 
and amplifies it via stimulated emission, and a disordered medium that partially traps 
light by multiple scattering. 
Lasing occurs when the loss due to absorption and leakage of light through open boundaries 
is compensated by light emission and amplification inside the medium. 
Stronger scattering increases the trap time of light and lowers the leakage loss. 
Letokhov discussed the diffusion process with gain that can lead to lasing with non-resonant 
feedback in the 1960s \cite{letokhov68a}. 
Early experiments, e.g., on dye solutions containing microparticles \cite{lawandy94},
showed a dramatic narrowing of the emission spectrum and a rapid increase of emission 
intensity at the frequencies around the maximal gain where the threshold 
condition is met. 
In contrast to the smooth and relatively broad lasing spectra, later experiments illustrated 
multiple sharp peaks of laser emission from semiconductor powder and disordered 
polymers \cite{cao99,frolov99a}. 
Those spectral peaks result from interference of scattered light in the random media. 
Although interference is not required for lasing action, it reduces light leakage at certain 
frequencies  \cite{milnerPRL05} and facilitates lasing by lowering the threshold (gain = loss). 
Thus lasing occurs at those frequencies, producing emission with high first-order coherence 
(narrow spectral width) and second-order coherence (suppression of photon number fluctuations in single
modes) \cite{cao01prl}. 
In addition to the reproducible lasing peaks, stochastic spikes were observed in single-shot 
emission spectra with pulsed excitation \cite{Mujumdar, mujumdar07, wu07OL, wu08}. 
These spikes are completely different in frequency from shot to shot, and are attributed to 
strong amplification of spontaneously emitted photons along long paths. 

The rich phenomena of random lasers have not been well understood so far. 
The diffusion model including gain can describe the narrowing of broad emission spectra 
\cite{letokhov68a}, but not the appearance of discrete lasing peaks since it neglects the 
interference effects. 
Semiclassical laser theory, based on the Maxwell's equations, can predict the lasing peaks  \cite{souk00,vannestePRL01} 
but not the stochastic emission spikes because it does not take into account the spontaneous emission.
The spectral width of lasing modes cannot be calculated either without spontaneous emission. 
We do not know how the laser linewidth compares to the frequency spacing of lasing modes. 
If the former is larger than the latter, the lasing peaks are indistinguishable no matter how 
fine the spectral resolution is. 
A diffusive or ballistic system, especially with higher dimensionality, contains a huge 
number of resonances that overlap spatially and spectrally. 
Although these resonances have similar lasing thresholds, they cannot all lase simultaneously
because of gain depletion \cite{caoPRB03,tureciSci}.
A large fluctuation of lasing spectra have been observed experimentally 
\cite{anglos04, Mujumdar, noginovLPL04, markushevLP05, sharmaOL06, mujumdar07, lagen07sts, markushevLP07, elDardiryPRA10}. 
The lasing modes are sensitive to small perturbations and noise, e.g., fluctuation of pump 
pulse energy, spatial variation of pump intensity, etc. \cite{tureciSci}. 
In addition to the extrinsic noise, the number of spontaneously emitted photons 
participating in the buildup of laser emission in any mode may fluctuate from shot to shot, 
leading to variations of lasing peak height \cite{markushevLP07,wiersman}. 
Such intrinsic fluctuations are missed by semiclassical laser theory. 
Since random laser thresholds are usually higher than conventional laser thresholds 
due to weaker optical confinement, stronger pumping is required, making the amplified 
spontaneous emission (ASE) stronger. 
Semiclassical laser theory, which neglects ASE, cannot capture the transition from the 
amplification of spontaneous emission to lasing oscillation that has been observed 
experimentally \cite{cao00}. 

Note that spontaneous emission only contributes to part of the intrinsic noise, which also 
includes the fluctuations induced by atomic dephasing, pumping, optical leakage, etc. 
Since intrinsic noise plays an essential role in random lasing behavior, it must be treated 
properly. 
There have been significant advances in theoretical studies on photon statistics of random 
lasers and amplifiers \cite{beenakker98, hacken01, patra02, florescu04,skipetrov09}. 
Most of them are based on full quantum treatments of noise in the modal description. 
For a random system, the mode structures are complex and unknown \textit{a priori}. 
Thus it is desirable to introduce noise without prior knowledge of modes. 
Some previous studies based on light diffusion and random walks  \cite{lepriPRA07} do not need mode information, 
but they ignore light interference that is essential to the formation of resonant lasing modes.  

In this paper we incorporate intrinsic noise into the numerical simulation of random lasers. 
The numerical method is based on the finite-difference time-domain (FDTD) formulation we 
recently developed to study the effects of noise on light-atom interaction in complex systems without prior 
knowledge of resonances \cite{andreasen08, andreasen09jlt}. 
The interference effects and the openness of the system are fully accounted for with
Maxwell's equations and absorbing boundary conditions \cite{tafl05}.   
The incorporation of the Bloch equations for the density of states of atoms simulate the 
dynamics of atoms and their interaction with light \cite{ziol95}. 
In the many-atom and many-photon limit the quantum fluctuations can be simulated by classical
noise terms \cite{drum91}. 
Based on the fluctuation-dissipation theorem, we consider noise associated with three 
dissipation mechanisms for atoms (described in detail in \cite{andreasen09jlt}) 
(i) dephasing events,
(ii) excited state decay,
(iii) incoherent pumping (from ground state to excited state).
Noise related to field decay is negligible because the photon energy at visible frequencies 
is much larger than the thermal energy at room temperature. 
At higher temperatures or longer wavelengths, where this noise becomes significant, it can be 
incorporated into the FDTD algorithm following the approach we developed in 
\cite{andreasen08}. 

Here we study the effects of intrinsic noise on the steady-state properties of random lasers 
in one dimension (1D). 
Results from systems of different scattering strengths are presented which probe 
varying degrees of light leakiness and spectral mode overlap.
For the first time, we are able to simulate the transition from ASE to lasing 
using the stochastic Maxwell-Bloch equations.
Stochastic emission spikes are reproduced with similar statistics to the experimental data 
reported previously \cite{wu07OL,wu08}. 
The spectral width of the broad ASE peak is calculated as a function of pumping rate. 
It displays a dramatic decrease with increasing pump level as seen experimentally 
\cite{lawandy94}. 
The linewidths of individual lasing modes are also computed and compared to the 
Schawlow-Townes linewidth of single mode lasers. 
A comparison of the results of simulations with noise to the simulations of the same active 
system without noise illustrates that noise effects are strongest in the transition regime 
from ASE to lasing. 

This paper is organized as follows. 
In Sec. \ref{sec:rsg} information of the random systems studied here is provided.
An analysis of resonances in these systems without gain is carried out in 
Sec. \ref{sec:passive}.
In Sec. \ref{sec:ran}, the FDTD formulation for the stochastic Maxwell-Bloch equations are given and some 
numerical issues related to noise are discussed.
Results of calculations using the Maxwell-Bloch equations both with and without 
noise are presented for random systems with spectrally overlapping resonances in Sec. \ref{sec:overlap} 
and with  non-overlapping resonances in Sec. \ref{sec:nonovlp}.
Our main conclusions are drawn in Sec. \ref{sec:conclusion}.

\section{Random Systems\label{sec:rsg}}

Two 1D random systems are considered here with different degrees of spectral overlap of resonances.
Both consist of $N=41$ layers. 
The dielectric layers with index of refraction $n_1>1$ alternate with air gaps 
($n_2=1$) resulting in a spatially modulated index of refraction $n(x)$. 
The scattering strength is varied by adjusting the index contrast $\Delta n=n_1/n_2-1$.
The system is randomized by specifying different thicknesses for each of the layers as 
$d_{1,2} = \left<d_{1,2}\right>(1+\eta\zeta)$, where
$\left<d_1\right>$ and $\left<d_2\right>$ are the average thicknesses of the layers, 
$0 < \eta < 1$ represents the degree of randomness, and $\zeta$ is a random number uniformly 
distributed in (-1,1). 
The average thicknesses are $\left< d_1 \right> = 100$ nm and $\left<d_2\right> = 200$ nm,  
giving a total average length of $\left<L\right> = 6100$ nm.
The grid origin $x=0$ is at the left boundary of the structure, and the length of the random 
structure $L$ is normalized to $\left<L\right>$.
The degree of randomness is set to $\eta = 0.9$ and the index of refraction outside the random 
media is $n_0 = 1$.

\begin{figure}
\begin{center}
\begin{tabular}{m{.5cm}m{7cm}}
  (a)\vspace{4cm} & \includegraphics[width=7cm]{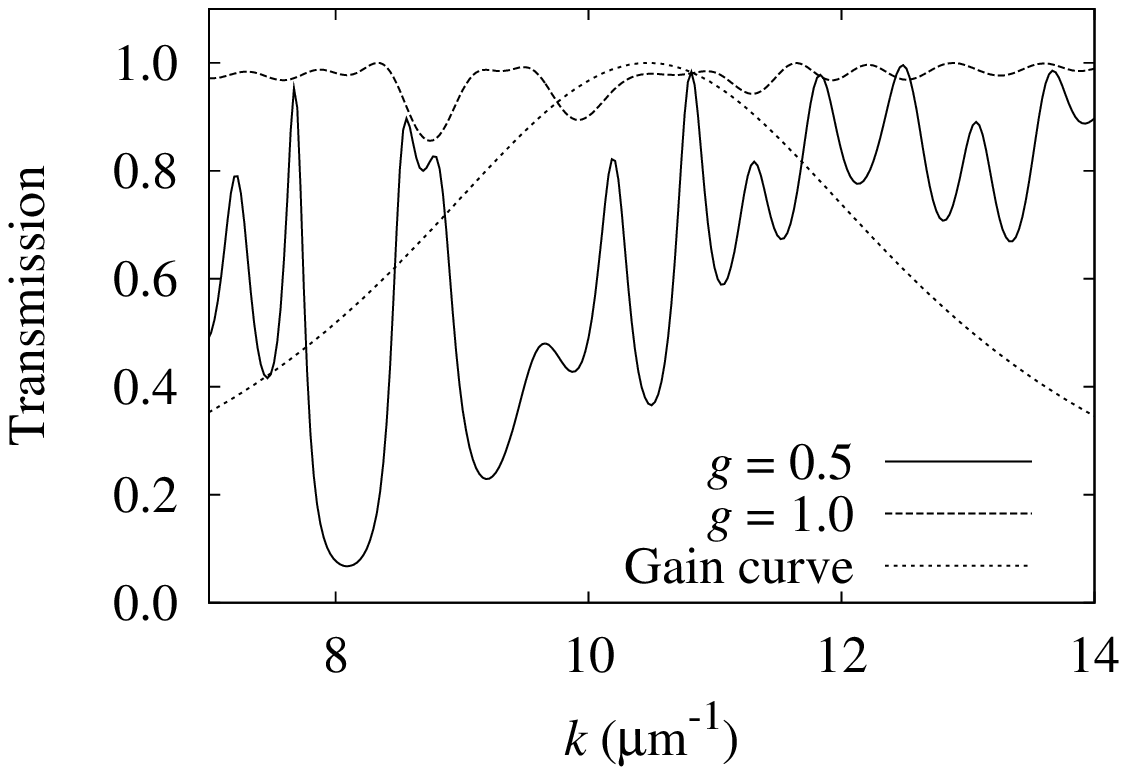}\\
  (b)\vspace{4cm} & \includegraphics[width=7cm]{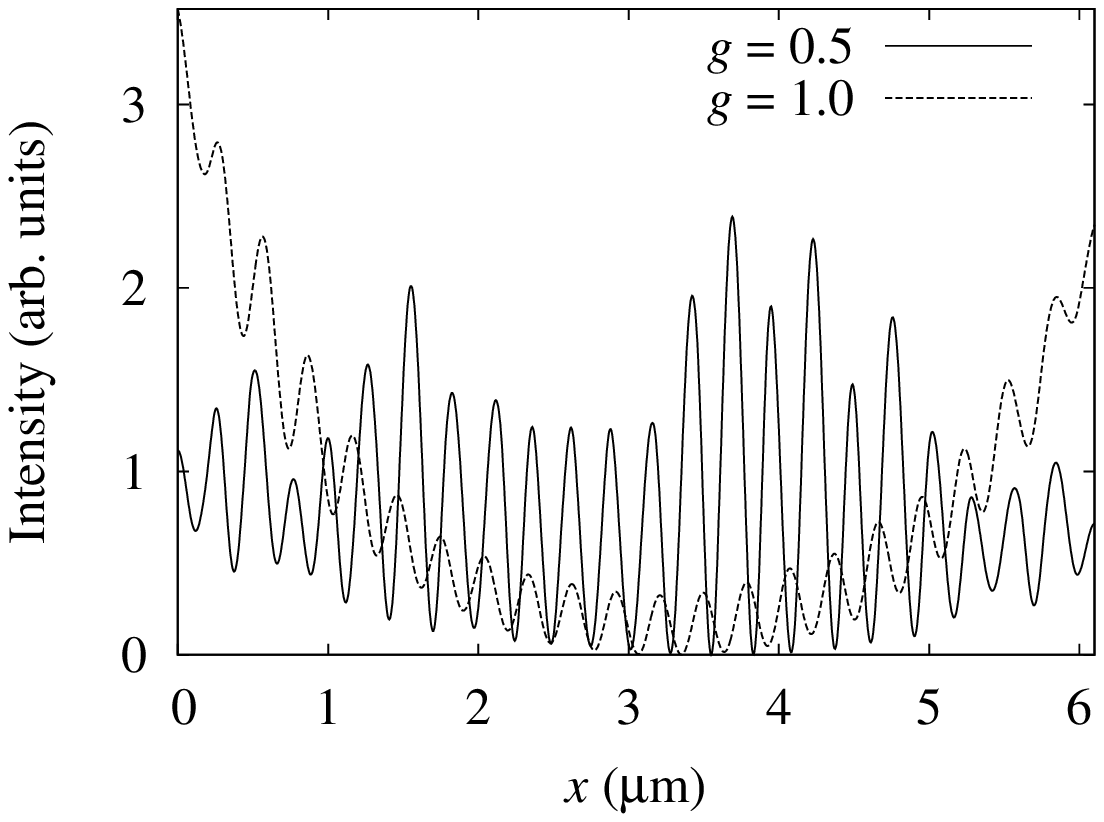}\\
\end{tabular}
  \caption{\label{fig:fig1}
    (a) Transmission spectra $\mathcal{T}(k)$ of passive random systems with $g=0.5$ (solid line) and
    $g=1.0$ (dashed line). The gain curve (dotted line) for the Maxwell-Bloch simulations
    is also shown.
    (b) Intensity distribution of a representative quasimode in a random system with
    $g=0.5$ (solid line) and $g=1.0$ (dashed line). 
  }
\end{center}
\end{figure}

The degree of mode overlap is adjusted by the refractive index $n_1$ of the dielectric layers.
The Thouless number $g$, which reveals the amount of spectral overlap of resonances of these 
random systems, is given by the ratio of the average resonance decay rate to the average 
frequency spacing $g=\left<k_i\right>/\left<\Delta k\right>$.
In the first case, $n_1 = 1.05$ ($\Delta n=0.05$), $g=1.0$ and the resonances overlap in 
frequency. 
In the second case, $n_1 = 1.25$ ($\Delta n=0.25$), $g=0.5$ and the resonances are fairly 
well separated. 
The localization length $\xi$ is obtained from the variation of transmission over the system 
length, and averaged over the wavelength range of interest (400 nm to 1200 nm). 
$\left<\xi\right> = 340$ $\mu$m for the first case, and $\left<\xi\right> = 13$ $\mu$m for the 
second one. 
Since $\left<\xi\right>$ is much larger than the system length $L$, both systems are far from 
localization threshold. 
Figure \ref{fig:fig1}(a) shows the transmission spectra $\mathcal{T}(k)$ of both systems.
Resonance peaks are clearly narrower and better separated for the second system with $g=0.5$.

\section{Resonances of the Passive System\label{sec:passive}}

We calculate the resonances of the two systems in the absence of gain or absorption using the 
transfer matrix method. 
Because the system is open, light can escape through the boundaries. 
To satisfy the conditions that there are only outgoing waves through the boundaries, the wave vectors must be complex numbers, $\tilde{k} = k + ik_i$.  
The real part $k$ corresponds to the mode frequency $\omega$, $k = \omega /c$, where $c$ is the 
speed of light in vacuum. 
The imaginary part $k_i <0$; its amplitude is proportional to the decay rate $\gamma$ of the 
mode due to light leakage. 
The resulting field distributions associated with the solutions for these boundary conditions are 
the quasimodes of the passive system. 
Figure \ref{fig:fig1}(b) plots the electric field intensity distributions of representative 
quasimodes in the two systems studied here.
With $g=0.5$ the spatial distribution of electric field intensity is more 
concentrated inside the system than with $g=1.0$ where intensity distribution is concentrated 
on the boundaries of the system.

\begin{figure}
  \includegraphics[width=8.5cm]{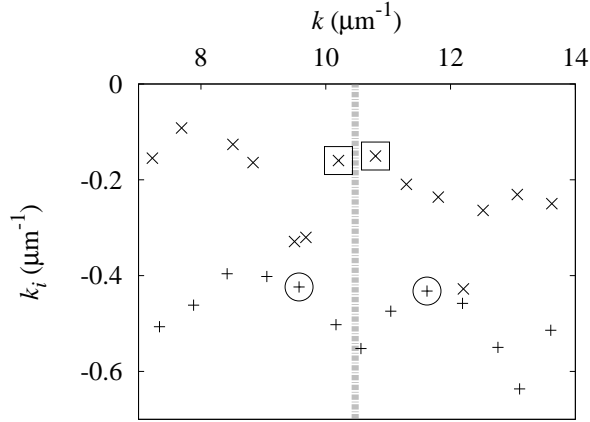}\\
  \caption{\label{fig:fig2}
    Frequencies $k$ and decay rates $k_i$ of quasimodes in two random systems with $g=1.0$ ($+$'s)
    and $g=0.5$ ($\times$'s).
    The vertical dashed gray line marks the atomic transition frequency $k_a$ in the Maxwell-Bloch simulations.
    The two strongest lasing modes found in the following 
    section are bounded by (circles) for $g=1.0$ and (squares) for $g=0.5$.
    The circled modes have smaller decay rates than the modes nearest $k_a$.
  }
\end{figure}

Figure \ref{fig:fig2} shows all quasimodes of the two systems in the complex-$\tilde{k}$ plane 
within the wavelength range of interest. 
For $g=1.0$ the separation of decay rates between neighboring modes is quite small, 
leading to significant spectral overlap of modes.
With gain included, the lasing thresholds of most modes should be very similar.
However, the simulations of lasing in the following sections include a frequency-dependent
gain curve centered at $k_a$ (shown as the vertical line in Fig. \ref{fig:fig2}). 
A balance of lower decay rate (smaller amplitude of $k_i$) and higher gain ($k$ closer to $k_a$) 
selects the modes that are amplified the most, e.g., the two modes circled in 
Fig. \ref{fig:fig2}. 

For the system with $g=0.5$ in Fig. \ref{fig:fig2}, the resonances have smaller 
amplitudes of $k_i$ and thus lower decay rates. 
This is a result of greater confinement of light due to the higher index contrast.
Furthermore, decay rates are more separated in general.
As suggested by the narrow peaks around $k=10.5$ $\mu$m$^{-1}$ in the transmission 
spectrum in Fig. \ref{fig:fig1}(a), the two modes nearest $k_a$ have relatively small decay 
rates.
They are also fairly well separated from neighboring modes. 
Thus, these two modes should have the lowest lasing thresholds.

\section{FDTD simulation of Stochastic Maxwell-Bloch equations\label{sec:ran}}

We consider two-level atoms uniformly distributed over the entire random system.
The atomic transition frequency is set to $k_a=10.5$ $\mu$m$^{-1}$, the corresponding wavelength 
$\lambda_a=600$ nm. 
The lifetime of atoms in the excited state $T_1$ and the dephasing time $T_2$ are included in the 
Bloch equations for the density of states of atoms.
The spectral width of the gain regime is given by $\Delta k_a=(1/T_1+2/T_2)/c$ \cite{siegbook}.
We set $T_1=1.0$ ps.
The values of $T_2$ are chosen such that the gain spectrum spans ten quasimodes of the passive 
system, i.e., $\Delta k_a=10\left<\Delta k\right>$.
The average frequency spacing $\left<\Delta k\right>$ is slightly different for the two cases studied.
For $g=0.5$, $T_2=1.4$ fs and $\Delta k_a=4.7$ $\mu$m$^{-1}$.
For $g=1.0$, $T_2=1.3$ fs and $\Delta k_a=5.0$ $\mu$m$^{-1}$,
as shown by the dotted line in Fig. \ref{fig:fig1}(a).
We also include incoherent pumping of atoms from level 1 to level 2.
The rate of atoms being pumped is proportional to the population of atoms in level 1 $\rho_{11}$, 
and the proportional coefficient $P_r$ is called the pumping rate.

To introduce noise to the Bloch equations, we used the stochastic $c$-number equations that are 
derived from the quantum Langevin equations in the many-atom limit \cite{drum91}. 
The noise sources in these equations are from both the dissipation of the system and the nonlinearity in 
the Hamiltonian. 
The latter represents the nonclassical component of noise, giving rise to nonclassical statistical 
behavior. 
Since we are interested in the classical behavior of macroscopic systems, such as ASE and lasing, 
we neglect the nonclassical noise in our simulation. 
The classical noise results from the decay, dephasing and pumping of atoms, as dictated by the 
fluctuation-dissipation theorem. 
The amplitude of classical noise accompanying the field decay is proportional to the square root of 
the thermal photon number.
At room temperature the number of thermal photons at visible frequencies is negligible, 
thus the noise related to the field decay is ignored here. 

The stochastic simulations solve for the atomic population of the excited states 
$\rho_{22}$ and the atomic polarization $\rho_1=\rho_{12} + \rho_{21}$ and 
$\rho_2=i(\rho_{12} - \rho_{21})$ which couple to Maxwell's equations.
The stochastic equations solved at each grid point in space are 
\begin{align}
  \frac{d\rho_1(x,t)}{dt} =& ck_a\rho_2(x,t) - \frac{1}{T_2}\rho_1(x,t) +
  \Gamma_1(x,t) \nonumber\\
  \frac{d\rho_2(x,t)}{dt} =& -ck_a\rho_1(x,t) +
  \frac{2|\gamma|}{\hbar}E_z(x,t)\left(2\rho_{22}(x,t)-N_s\right)\nonumber\\
  &- \frac{1}{T_2}\rho_2(x,t) +   \Gamma_2(x,t) \nonumber\\
  \frac{d\rho_{22}(x,t)}{dt} =& -\frac{|\gamma|}{\hbar}E_z(x,t)\rho_2(x,t)
  - \frac{1}{T_1}\rho_{22}(x,t)
  \nonumber\\
  &+ \frac{P_r}{T_1}(N_s-\rho_{22}(x,t)) + \Gamma_{22}(x,t), \label{eq:theory}
\end{align}
where $E_z$ is the electric field, $\gamma$ is the dipole coupling term, 
$N_s$ is the number of atoms per grid cell, and the noise terms
\begin{align}
  \Gamma_1(x,t) =& 2\xi_1(t)\sqrt{\gamma_p\rho_{22}(x,t)}\nonumber\\
  \Gamma_2(x,t) =& -2\xi_2(t)\sqrt{\gamma_p\rho_{22}(x,t)}\nonumber\\
  \Gamma_{22}(x,t) =& \xi_3(t)\sqrt{\rho_{22}(x,t)/T_1+ P_r \rho_{11}(x,t)/T_1},\label{eq:Gamma122122}
\end{align}
where $\gamma_p=1/T_2-1/2T_1$. 
The $\xi_j$ terms are real, Gaussian, random variables with zero mean and the
following correlation relation
\begin{equation}
  \left< \xi_j(t)\xi_k(t')\right> = \delta_{jk}\delta(t-t'),
\end{equation}
where $j, k = 1, 2, 3$.
We assume $T_2 \ll T_1$, and the pump fluctuations in $\Gamma_1$ and $\Gamma_2$ are neglected 
because they are orders of magnitude smaller than the noise due to dephasing. 
The resulting Maxwell-Bloch (MB) equations are solved through a parallel FDTD 
implementation with the spatial grid step $\Delta x = 1.0$ nm and the temporal 
step $\Delta t = 1.7\times 10^{-18}$ s.
In cases where noise is not included, the system is excited by a Gaussian-sinusoidal pulse of 
center frequency $k_0=k_a$ and spectral width $\Delta k_0=\Delta k_a$.

An issue concerning the simulation with noise arises when $\rho_{11}$ or $\rho_{22}$ is close to zero. 
To keep the atomic populations in both levels positive for a large range of pumping rates, 
we set the system initially at the transparency point, i.e., $\rho_3(t=0)=\rho_{22}-\rho_{11}=0$. 
Moreover, we assume the atomic density is large, $N_{atom}/V = 4.3 \times 10^{17}$ cm$^{-3}$ 
\cite{andreasendiss}.
Small variations to the initial population do not affect the final steady-state results.
Furthermore, the high frequency components of the noise excite the modes resonating within single 
air gaps sandwiched between dielectric layers of index $n_1 > 1$.
These high frequency contributions are ignored completely by considering only the 
electromagnetic fields within the wavelength range 400 nm $< \lambda < 1200$ nm.

With noise terms included in the Maxwell-Bloch equations, all quantities fluctuate in time. 
Eventually their values averaged over small time windows are nearly constant.
By comparing the spectra of output light taken over different temporal ranges up to $t=267$ ps, 
we find a steady state is reached by 16.6 ps for all pumping rates considered here.
Hence, the output spectra obtained after 16.6 ps represent the steady state behavior.
The output field is sampled at the grid point $x=L$ at the right boundary of the random system. 
The results from this point are identical in character to those from any point outside the system and 
before the absorbing boundary layer. 

\section{ASE and Lasing in a System with Overlapping Resonances\label{sec:overlap}}

\subsection{Input-output relation}

Starting from the random system with $g=1.0$, we investigate the transition from spontaneous emission 
to ASE and to laser emission by examining the dependence of the steady-state output intensity $I_o$ 
on the pumping rate $P_r$.
To avoid erroneous contributions from high frequency components (mentioned in 
Sec. \ref{sec:rsg}), $I_o$ is found by a spectral integration
\begin{equation}
  I_o = \int_{k_l}^{k_u} |E(k')|^2 dk',\label{eq:Ik}
\end{equation}
where $k_l= k_a - \Delta k_a = 2\pi/1.2$ $\mu$m$^{-1}$ and 
$k_u = k_a + \Delta k_a =2\pi/0.4$ $\mu$m$^{-1}$.

\begin{figure}
  \includegraphics[width=8.5cm]{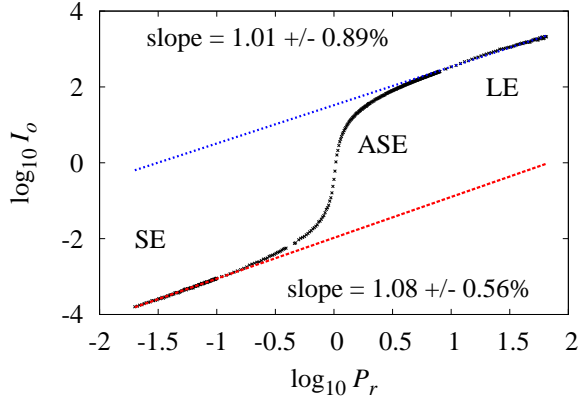}\\
  \caption{\label{fig:fig3} (Color online)
    Average steady-state emission intensity $I_o$ (black $\times$'s) vs. pumping rate 
    $P_r$ for a random system with $g=1.0$.
    Both $I_o$ and $P_r$ are plotted on log$_{10}$ scales to clearly show the regions of spontaneous emission (SE), 
    amplified spontaneous emission (ASE), and laser emission (LE).
    Red dashed line and blue dotted line are linear fits to the intensities of SE (with nearly constant reabsorption at $P_r < 0.1$) and LE 
    (well above the lasing threshold at $P_r > 10$), respectively. 
    The slopes, written next to the lines, are equal to one, reflecting linear increase of $I_o$ with $P_r$.  
    The intensity of ASE increases superlinearly with $(P_r)^p$.
  }
\end{figure}

Figure \ref{fig:fig3} plots $\log I_o$ versus $\log P_r$. 
The total pumping rate is normalized such that at $P_r = 1$ ($\log_{10}P_r=0$), the system without 
noise reaches the transparency point in the steady state ($\left<\rho_3(x)\right>_x=0$). 
With noise, $P_r = 1$ is just below the transparency point ($\left<\rho_3(x)\right>_x \lesssim 0$),
because the atomic population in level 2 ($\rho_{22}$) is reduced by spontaneous emission. 
The spontaneous emission intensity is linearly proportional to $\rho_{22}$. 
However, when $\rho_{22}< \rho_{11}$, the spontaneously emitted light can be reabsorbed. 
The amount of reabsorption is determined by $\rho_{11} - \rho_{22}$, which varies with $P_r$. 
At very low pumping ($P_r < 0.1$), $\rho_{11} \gg \rho_{22}$, thus $\rho_{11} - \rho_{22} \simeq 1$, 
and the amount of reabsorption is almost constant. 
As $\rho_{22}$ increases linearly with $P_r$, the output intensity of spontaneous emission grows 
linearly with $P_r$. 
At higher pumping $0.1 < P_r < 1$, the decrease of $\rho_{11}$ leads to a significant reduction in 
reabsorption. 
In fact the amount of reabsorption decreases nonlinearly with $P_r$, resulting in a superlinear 
increase of $I_o$ with $P_r$. 
Once the pumping rate is large enough to induce a population inversion  ($\rho_{11} < \rho_{22}$), 
the spontaneously emitted light experiences a net amplification. 
The ASE intensity increases superlinearly with $P_r$, as seen in Fig. \ref{fig:fig3}.
Even with the existence of population inversion, the rate of light amplification 
may be less than the leakage rate, and there is no lasing oscillation. 
Once the pump exceeds a threshold, light leakage is compensated by amplification, and lasing oscillation 
occurs. 
Well above the lasing threshold, the optical gain is saturated and the growth of $I_o$ with $P_r$ becomes linear again.

To ensure these results are not limited to the particular configuration considered here, 
the simulations are repeated with another random seed (to initialize the noise terms) and 
another realization of a random structure with the same $g$.
The results are qualitatively similar.
Slight differences arise due to stochasticity.

\subsection{ASE spectra}

\begin{figure}
\begin{center}
  \includegraphics[width=4.25cm]{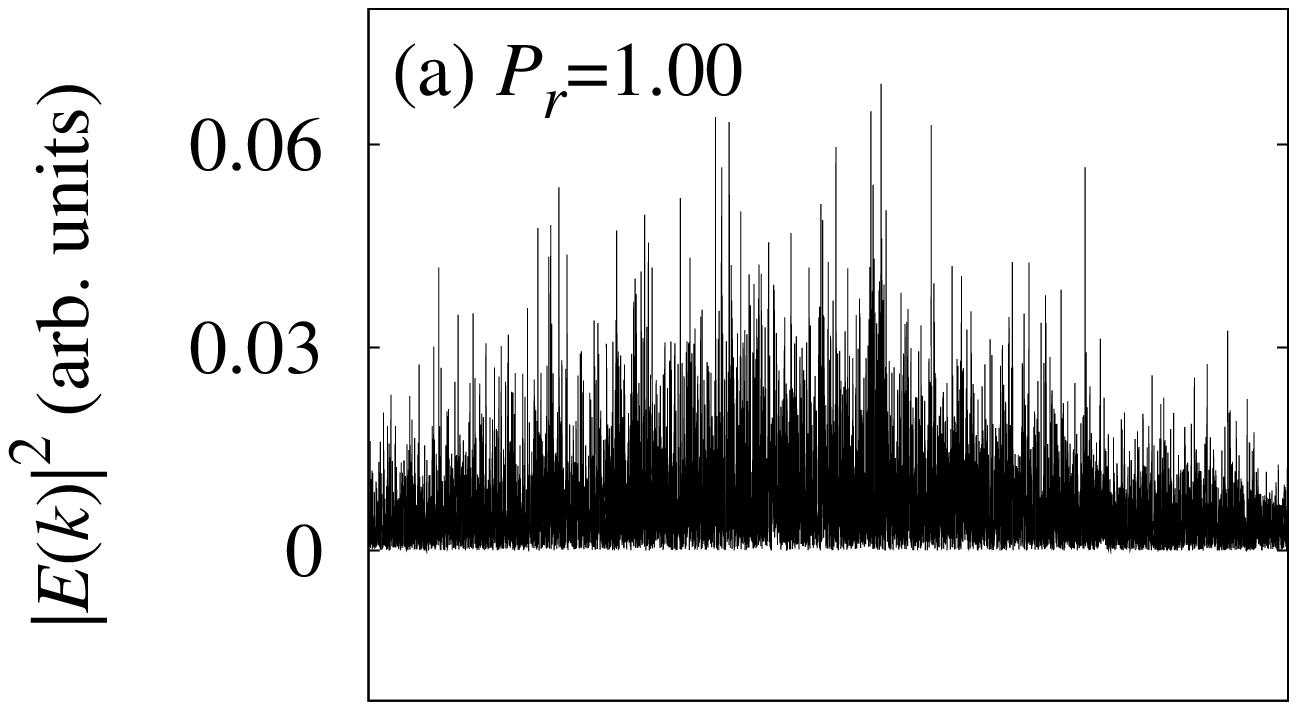}
  \includegraphics[width=4.25cm]{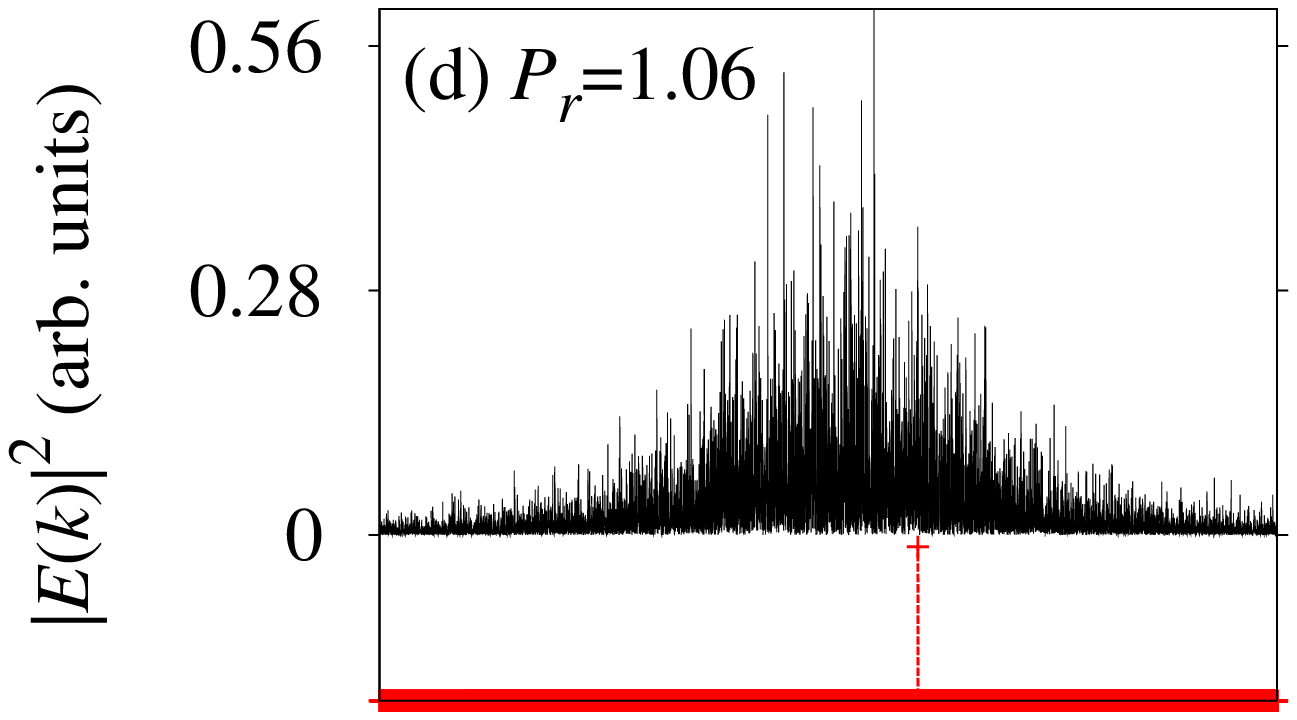}\\
  \includegraphics[width=4.25cm]{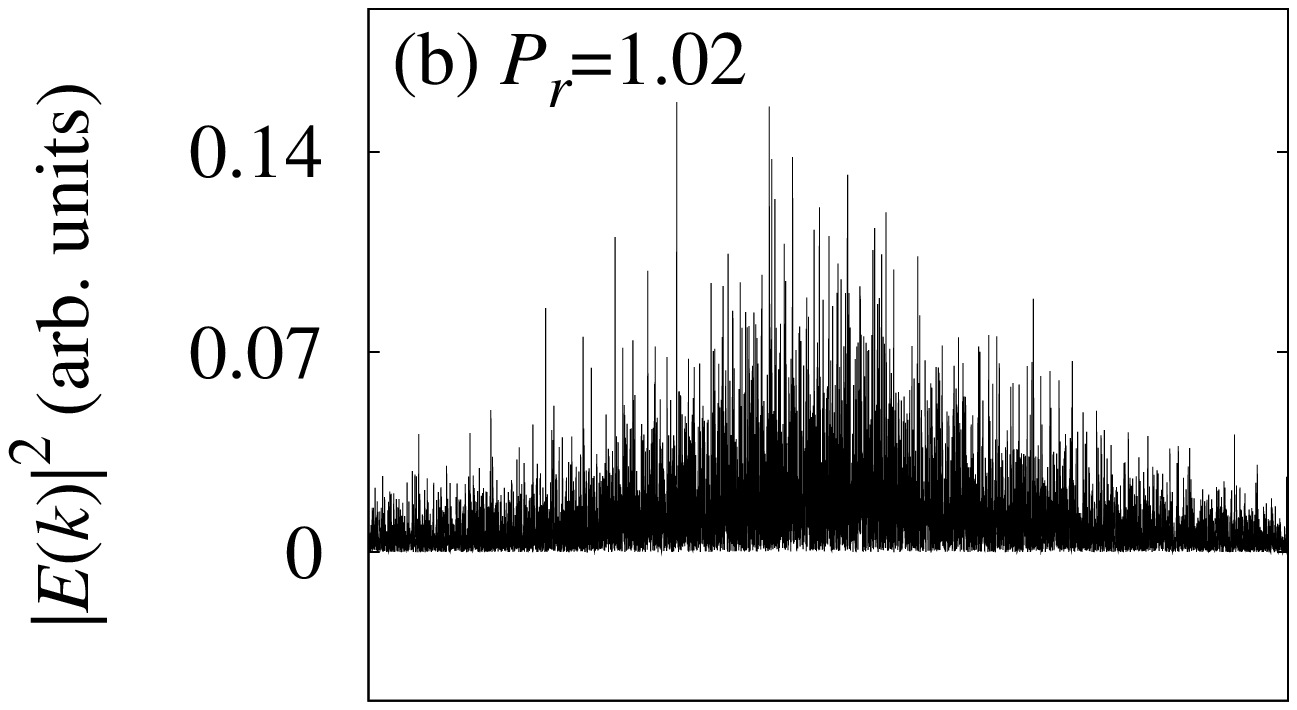}
  \includegraphics[width=4.25cm]{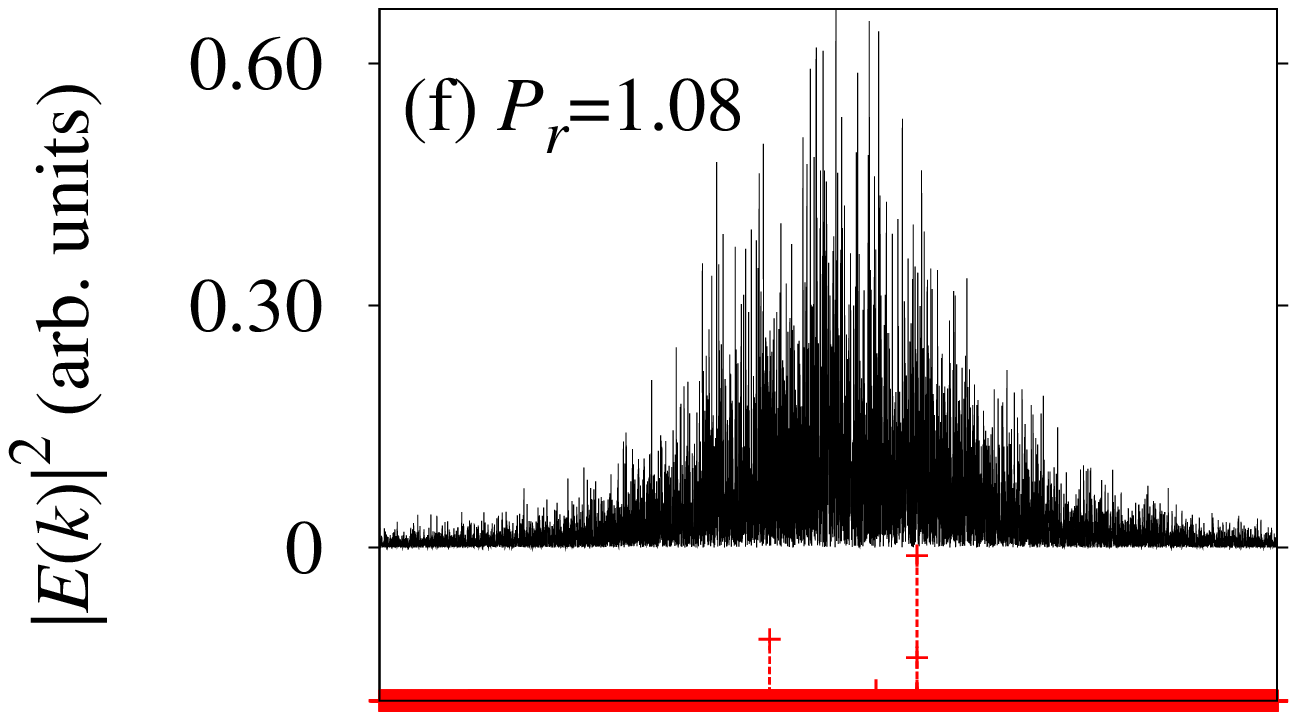}\\
  \includegraphics[width=4.25cm]{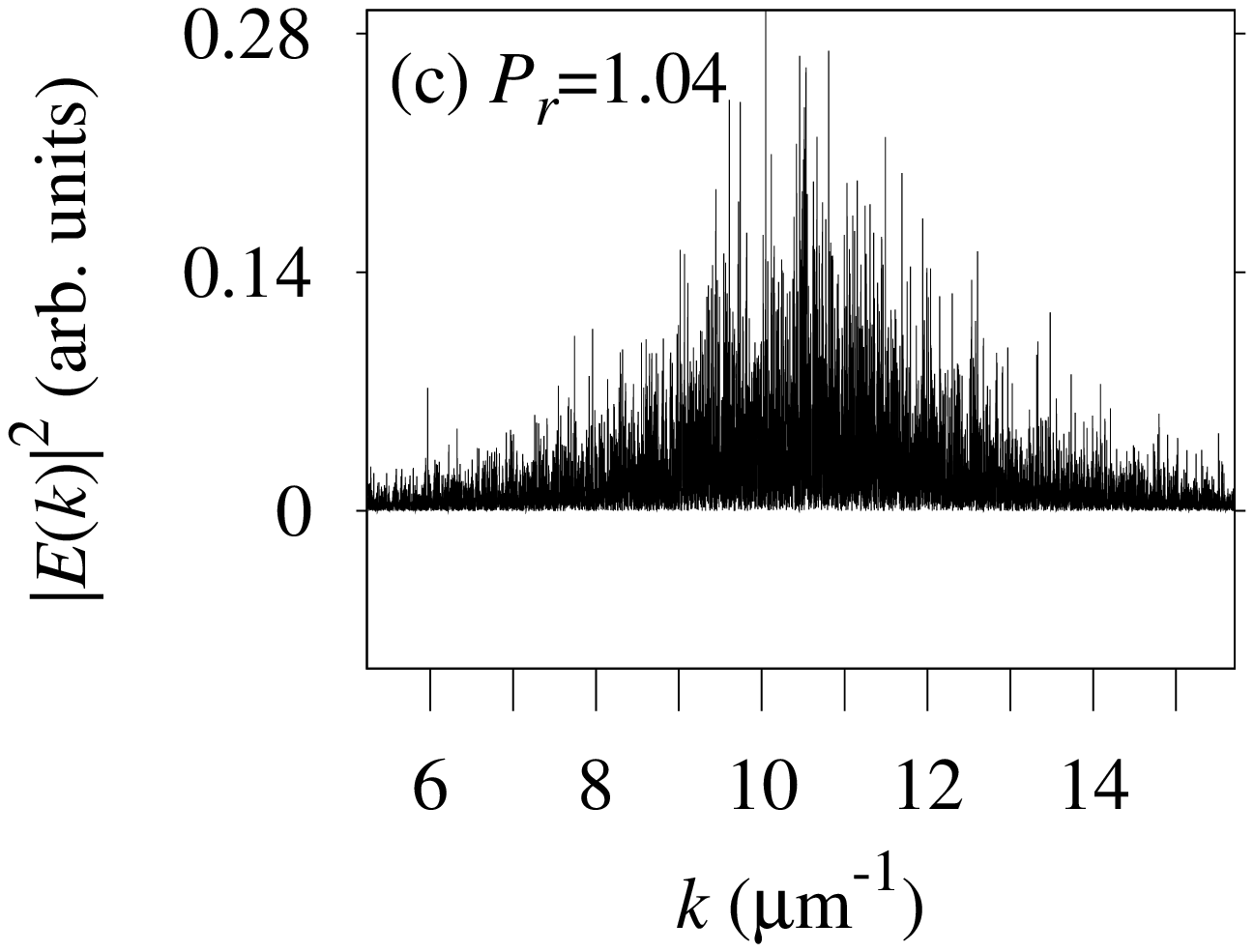}
  \includegraphics[width=4.25cm]{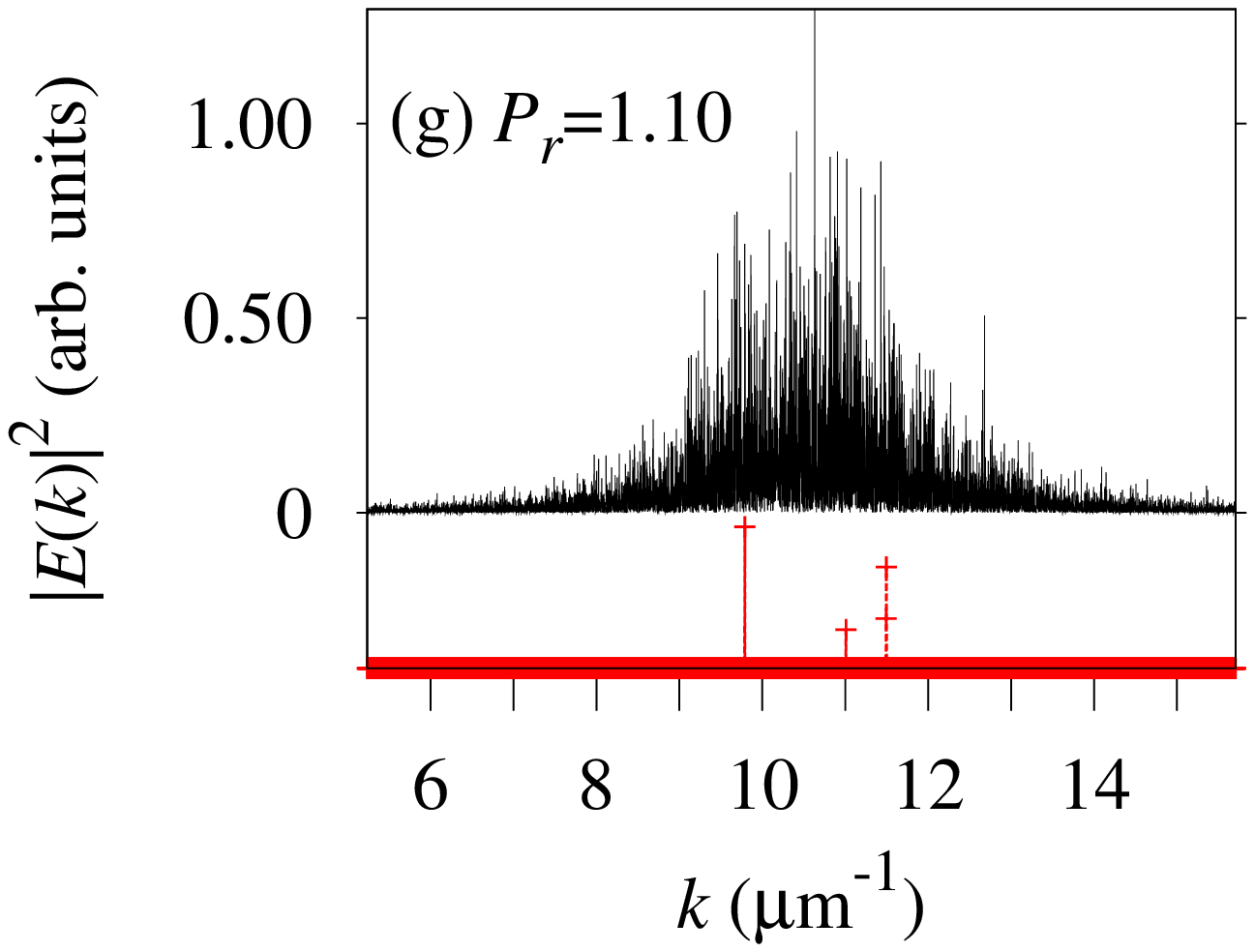}\\
  \caption{\label{fig:fig4} (Color online)
    Steady-state emission spectra $|E(k)|^2$ with noise (upper black line in each panel) compared to those 
    without noise (lower red line and crosses in the same panel) at the same pumping rate $P_r$ for a random system with $g=1.0$. 
    The values of $P_r$ are written in each panel. 
  }
\end{center}
\end{figure}

We Fourier-transform the output fields to obtain the emission spectra. 
Figure \ref{fig:fig4} shows the steady-state emission spectra with noise $|E(k)|^2$ in comparison to 
those without noise for increasing pumping rates.
At $P_r=1.00$ [Fig. \ref{fig:fig4}(a)], there is no net gain.
Without noise, the initial seed pulse dies away, and there is no signal at the steady state. 
With noise, the steady-state emission spectrum has a broad peak. 
It is centered at the atomic transition frequency $k_a=10.5$ $\mu$m$^{-1}$, 
resembling the spontaneous emission spectrum.  
On top of it there are many fine spikes whose frequencies change chaotically from one time window of 
Fourier transform to the next. 
They result from the stochastic emission process with their spectral width determined by the temporal 
length of the Fourier transform. 
Above the transparency point at $P_r=1.02$ [Fig. \ref{fig:fig4}(b)], 
the broad emission peak grows and narrows spectrally. 
This behavior is typical of ASE. 
Since the optical gain is frequency dependent, the emission intensity closer to $k_a$ is amplified more 
than that away from $k_a$, leading to a spectral narrowing. 
The stochastic emission spikes are also amplified, especially those closer to $k_a$ in frequency.  
As the pumping rate increases more [e.g., $P_r=1.04$ in Fig. \ref{fig:fig4}(c)], 
the broad peak grows and narrows further. 
Without noise, the emission spectra are blank, since ASE is neglected. 
When $P_r=1.06$ [Fig. \ref{fig:fig4}(d)], a single peak appears in the emission spectrum without noise.
This peak is a delta function with its ``linewidth'' merely determined by the integration time of the 
Fourier transformation. 
It shows lasing occurs in a single mode, which corresponds to the resonance of the passive system at 
$k=11.6$ $\mu$m$^{-1}$ in Fig. \ref{fig:fig2}. 
The lasing frequency is pulled towards $k_a$ at which gain is maximal. 
A further increase of $P_r$ to 1.08 leads to lasing in a second mode that corresponds to the resonance 
at $k=9.6$ $\mu$m$^{-1}$. 
Frequency pulling is also seen here. 
In the absence of noise, single mode lasing can be achieved by carefully adjusting $P_r$. 
This is no longer the case when noise is introduced. 
The emission spectra with noise look very different. 
There is clearly no single-mode lasing at any pumping rate. 
Intensity of the broad emission spectrum is modulated, as seen in Figs. \ref{fig:fig4}(d)--\ref{fig:fig4}(g).  
The emission intensities are enhanced not only at the frequencies of lasing peaks without noise, 
but also at some other frequencies.  
Optical amplification is stronger at the resonant frequencies of the system
and narrows the resonance peaks that overlap spectrally without gain. 
Reduced overlap of resonance peaks with gain results in a spectral modulation of emission intensity.  
For $P_r=1.10$ [Fig. \ref{fig:fig4}(g)], it is possible to associate the three lasing peaks for the 
case without noise to resonance peaks with noise. 
Additional resonance peaks are also discernible for the case with noise. 
Unlike the stochastic emission spikes, the frequencies of resonance peaks are stable in time, 
although their heights may vary from one time window of Fourier transform to the next. 
Their spectral widths are notably larger than those of the stochastic emission spikes. 

\begin{figure}
  \includegraphics[width=8.5cm]{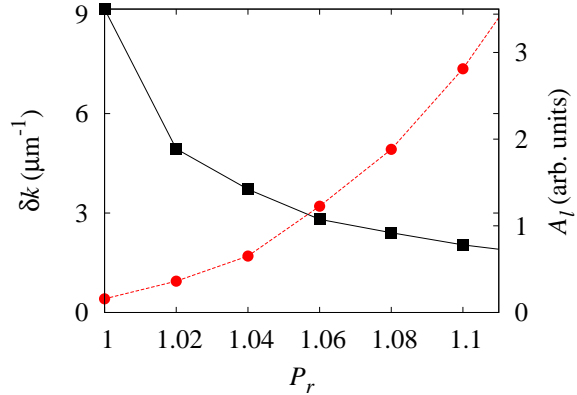}\\
  \caption{\label{fig:fig5} (Color online)
    Spectral width $\delta k$ (black squares) and amplitude $A_l$ (red circles) of the broad emission peak vs. 
    pumping rate $P_r$.
  }
\end{figure}

We extract the width of the broad emission spectrum in Fig. \ref{fig:fig4} at different pumping rates.
Because of the noisiness of the spectrum, we use a ``Lorentz error function'' to objectively obtain 
the spectral width.
A Lorentzian function $L(k)$ describes the spectrum 
\begin{equation}
  L(k) = \left(\frac{2A_l}{\pi}\right)\frac{(\delta k/2)^2}{(k-k_0)^2+(\delta k/2)^2},\label{eq:lor}
\end{equation}
where $A_l$ is the amplitude, $k_0$ is the center frequency, and $\delta k$ 
is the full-width at half-maximum (FWHM).
The Lorentz error function, given by
\begin{align}
  L_{EF}(k) &\equiv \int_{k_0}^{k} L(k') dk' \nonumber\\
  &= \left(\frac{A_l \delta k}{\pi}\right)\tan^{-1}\left(\frac{2(k-k_0)}{\delta k}\right),\label{eq:lef}
\end{align}
is used to fit the numerical data. 
The frequency range of the integration over $k'$ is limited to 
$k_l= k_a - \Delta k_a = 2\pi/1.2$ $\mu$m$^{-1}$ and $k_u= k_a + \Delta k_a=2\pi/0.4$ $\mu$m$^{-1}$.
Because of the preferential amplification of light with frequencies closer to
$k_a$, the emission spectrum is narrowed around $k_a$.
Thus, the center frequency is fixed at $k_0=k_a$ and not adjusted during the
fitting \cite{fityk} of Eq. (\ref{eq:lef}) to the data.
$A_l$ and $\delta k$ are the fitting parameters.

The values of $\delta k$ and $A_l$ obtained from the fitting is plotted against the pumping rate in 
Fig. \ref{fig:fig5}.
As $P_r$ increases from 1.00 to 1.10, $\delta k$ first decreases rapidly then decreases more slowly.
$A_l$ displays a superlinear increase with $P_r$.
Amplification allows the emission intensity to build up quickly around the atomic transition frequency. 
This increase results in a rapid narrowing of the emission spectrum.
Such behavior has been seen experimentally \cite{lawandy94}. 
Recent studies reveal that the spectral narrowing resembles a condensation process,  
as it can be predicted by a nonlinear differential equation identical to that governing the ultracold 
atoms \cite{conti08,conti10}. 

\begin{figure}
\begin{center}
\begin{tabular}{m{.5cm}m{7cm}}
  (a)\vspace{4cm} & \includegraphics[width=7cm]{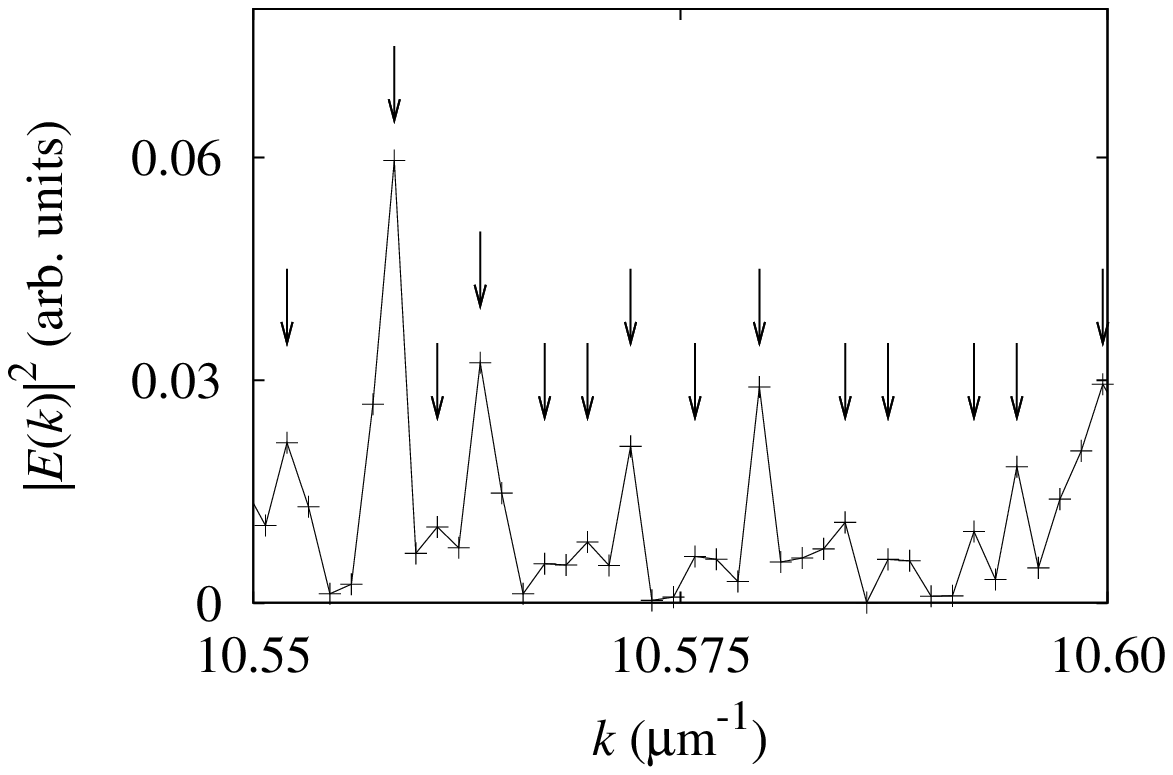}\\
  (b)\vspace{4cm} & \includegraphics[width=7cm]{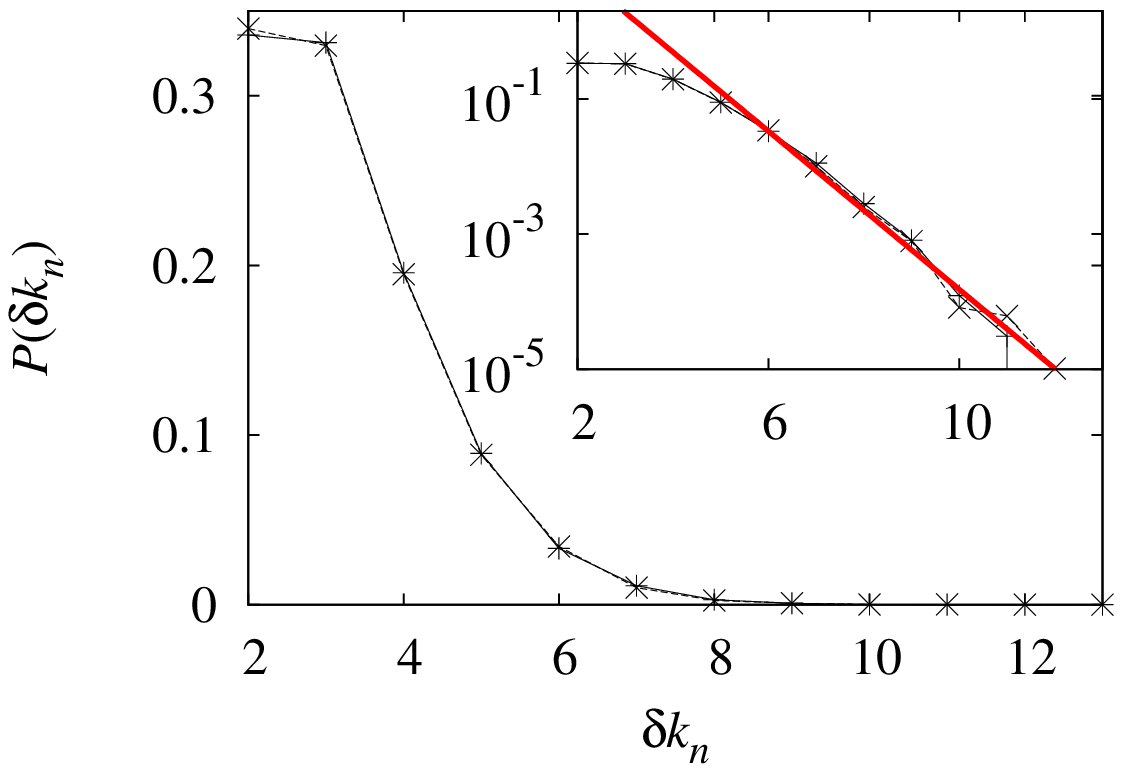}\\
\end{tabular}
  \caption{\label{fig:fig6} (Color online)
    (a) Steady-state emission spectra $|E(k)|^2$ with noise over a small frequency range
    illustrating stochastic emission spikes. 
    Crosses mark the wavelengths at which the emission intensities are obtained by the Fourier transform.
    Arrows mark spikes identified by a three-point peak finding method.   
    (b) Statistical distribution of frequency spacing of spikes
    $P(\delta k_n)$ at $P_r = 1.00$ ($+$'s) and $P_r = 2.00$ ($\times$'s), plotted on the linear scale (main panel) and logarithmic scale (inset). 
    An exponential fit is marked in the inset by a straight red line.
  }
\end{center}
\end{figure}

The stochastic emission spikes have also been observed in the ASE spectra experimentally 
\cite{Mujumdar, mujumdar07, wu07OL, wu08}. 
They are attributed to single spontaneous emission events which happen to take long open paths inside 
the amplifying random medium and pick up large gain. 
The emergence of these spikes does not rely on resonant feedback or coherent interference.
Their spectral width is determined by the temporal duration of the emission pulse. 
In principle, our classical noise model does not account for spontaneous emission on the single photon 
level. 
However, millions of photons are emitted and amplified in the macroscopic random media with gain.
Thus the quantum nature of photons can be ignored. 
We found the stochastic spikes in the emission spectra of our simulation bear similar characteristics
to the ASE spikes measured experimentally. 
An example of the stochastic spikes is shown in Fig. \ref{fig:fig6}(a), which is an enlargement
of the emission spectrum in Fig. \ref{fig:fig4}(a).
We find that the spectral width of the stochastic spikes is determined by the integration time of 
the Fourier transformation $T_F$.
Because of the long $T_F$, the spikes are usually much narrower than the lasing peaks, as long as the pumping rate is not too high.
In the emission spectrum the intensity is calculated at the frequency step  $\delta k_n$ that is determined by $T_F$. 
If the intensity at $k$ is larger than those at $k \pm \delta k_n$, a spike is found at $k$. 
Using this three-point peak-finding method, we extract the frequencies of spikes from the calculated 
spectra as shown in Fig. \ref{fig:fig6}(a), and compute the frequency spacing of adjacent spikes. 
Figure \ref{fig:fig6}(b) plots the statistical distribution $P(\delta k_n)$ 
of frequency spacing between adjacent stochastic spikes for $P_r = 1.00$ and $P_r = 2.00$. 
The two distributions coincide, revealing $P(\delta k_n)$ is independent of $P_r$. 
As evident from the log-linear plot in the inset of Fig. \ref{fig:fig6}(b), $P(\delta k_n)$ 
decays exponentially at large $\delta k_n$. 
This behavior is identical in character to the experimental result \cite{wu07OL, wu08}. 
Note that the level-off of $P(\delta k_n)$ at smaller $\delta k_n$ is an artifact of limited spectral 
resolution. 

\subsection{Lasing modes}

\begin{figure}
\begin{center}
  \includegraphics[width=4.25cm]{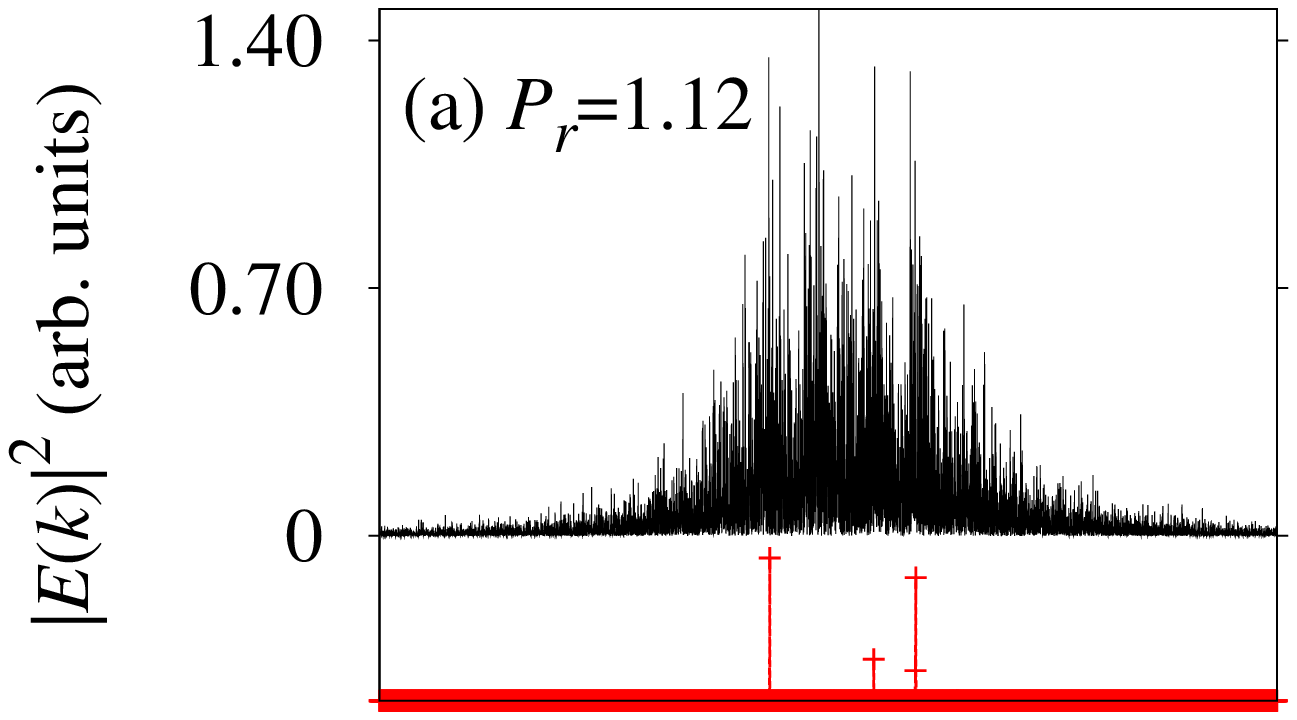}
  \includegraphics[width=4.25cm]{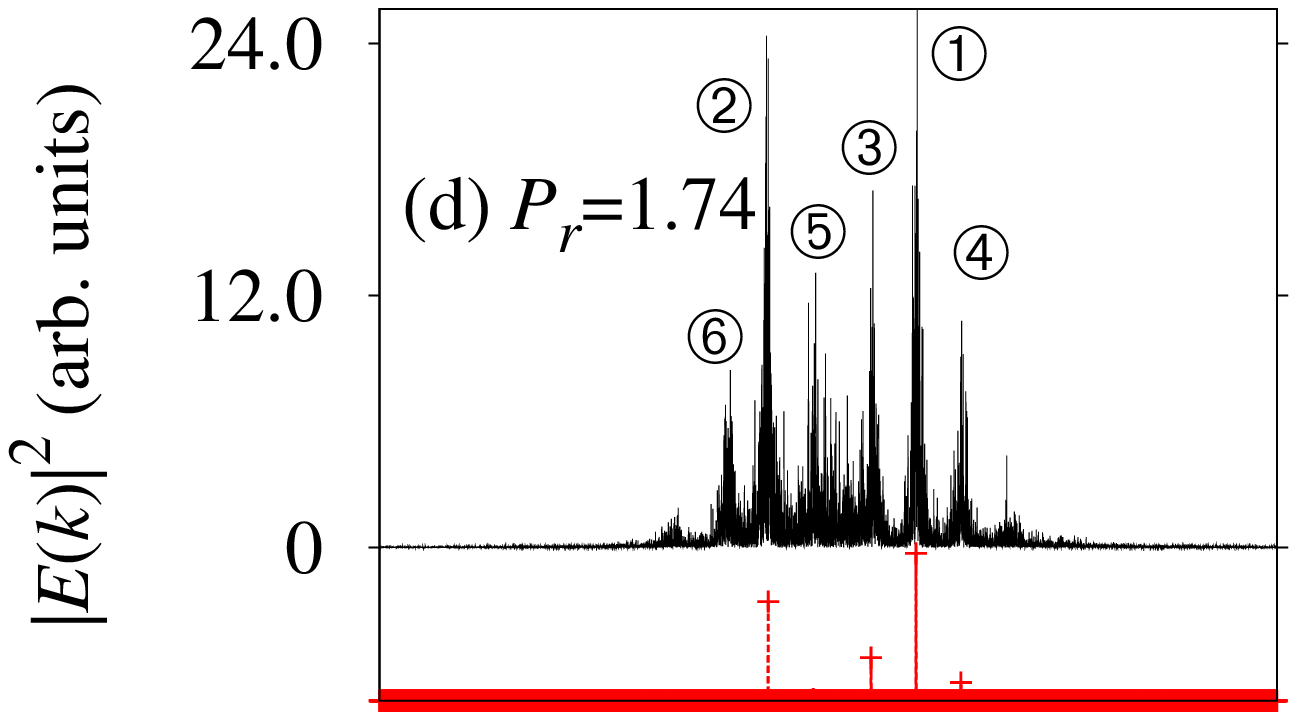}\\
  \includegraphics[width=4.25cm]{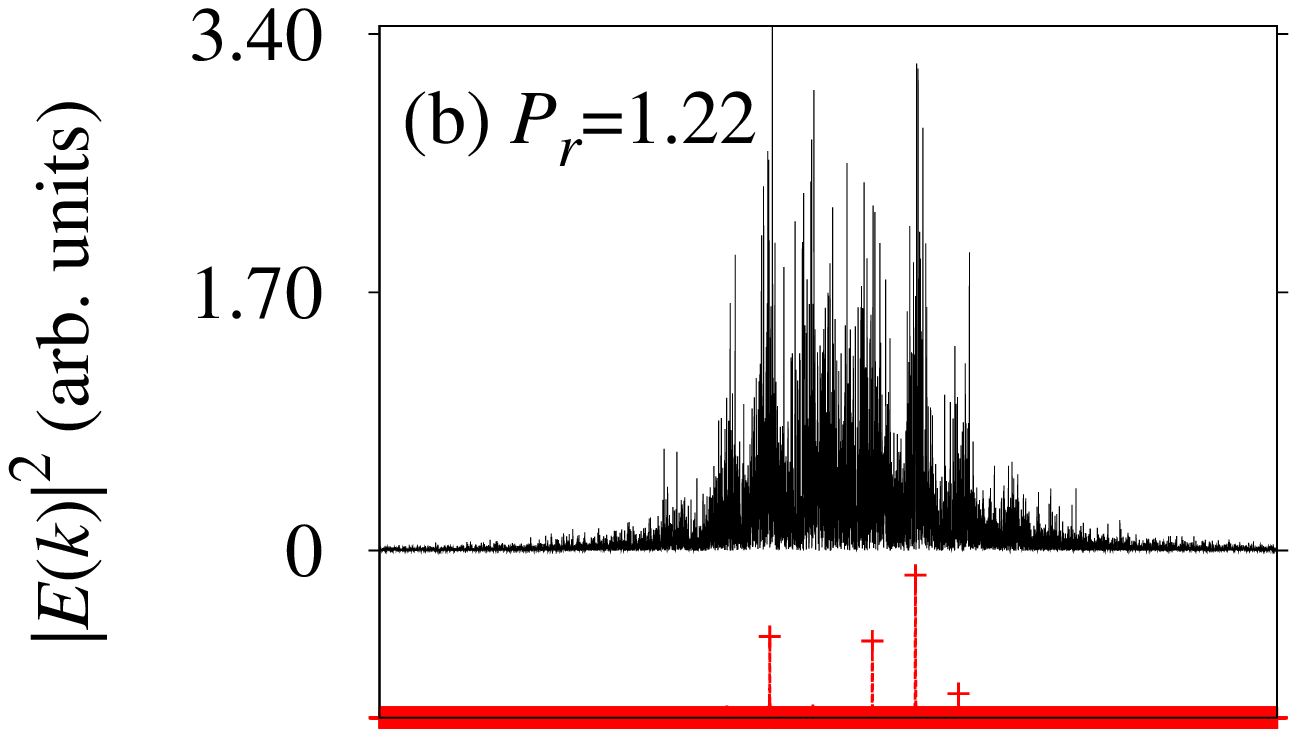}
  \includegraphics[width=4.25cm]{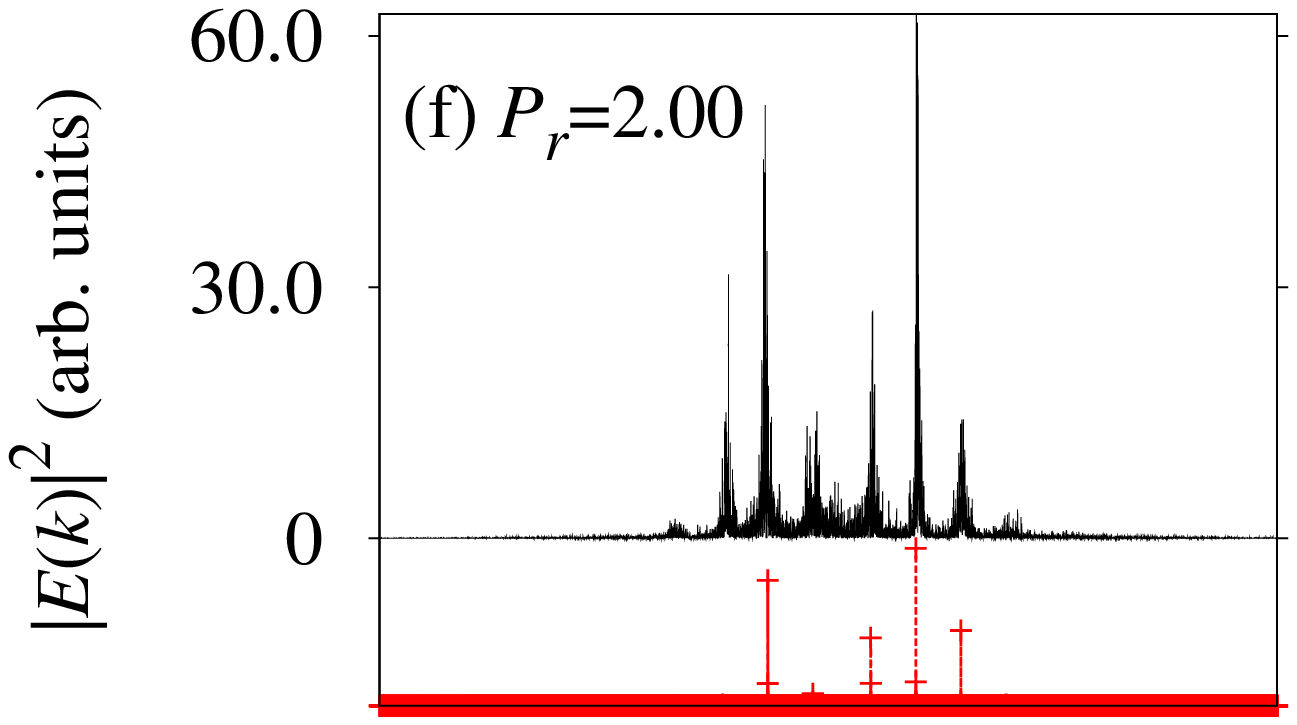}\\
  \includegraphics[width=4.25cm]{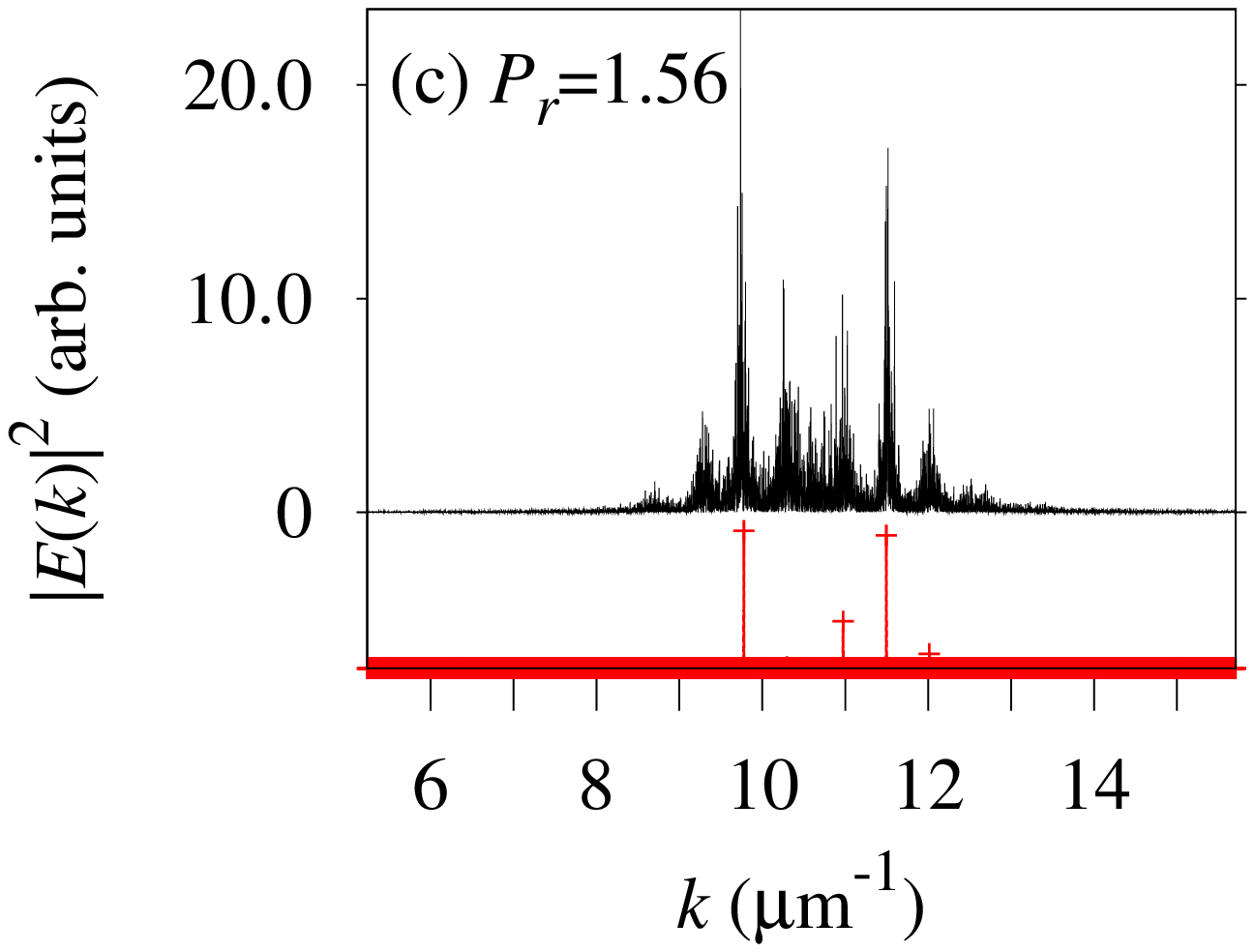}
  \includegraphics[width=4.25cm]{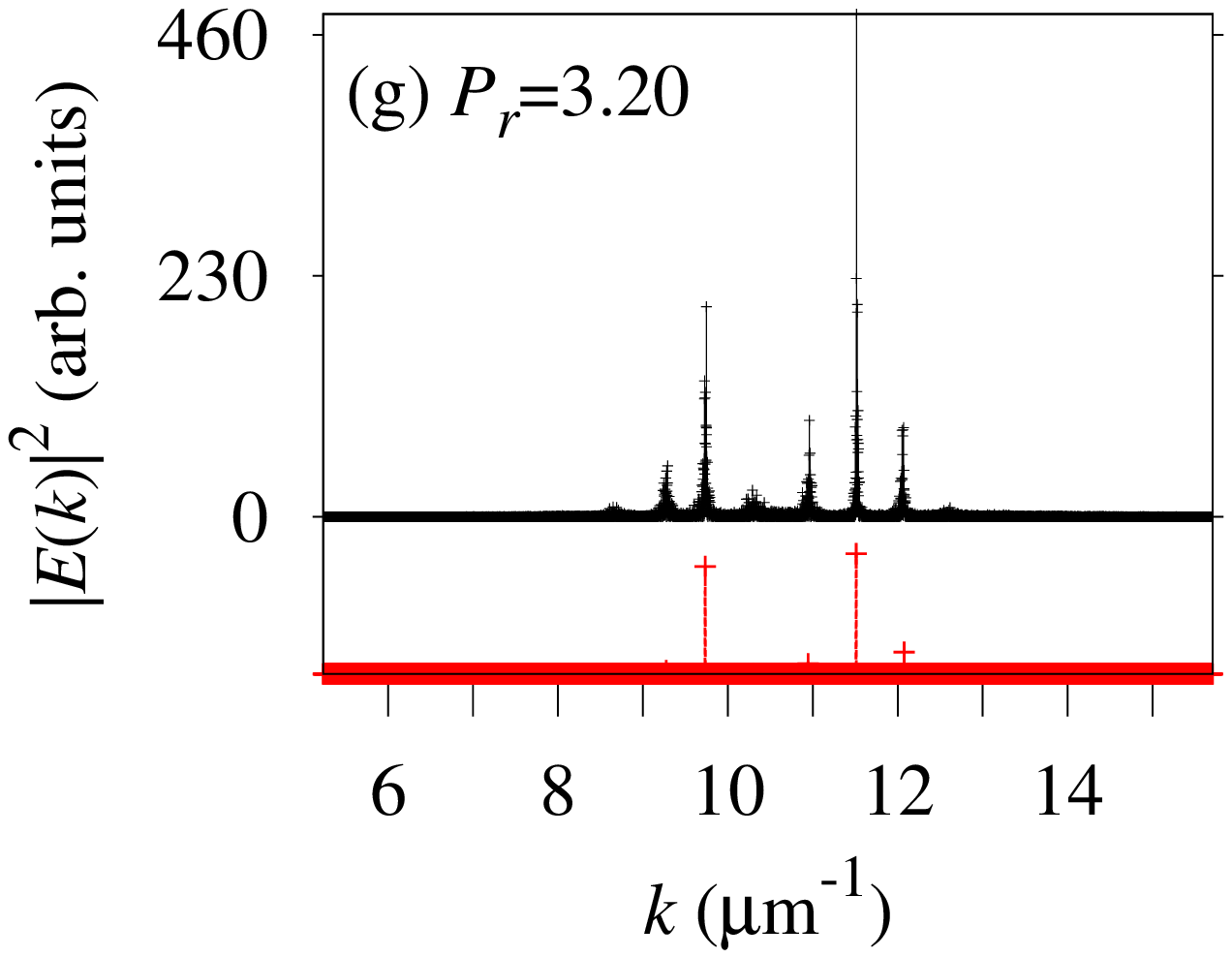}\\
  \caption{\label{fig:fig7} (Color online)
    Steady-state emission spectra $|E(k)|^2$ with noise (upper black line in each panel) compared to those 
    without noise (lower red line and crosses in the same panel) at the same pumping rate $P_r$ for a random system with $g=1.0$. 
    The values of $P_r$ are written in each panel.
  }
\end{center}
\end{figure}

The resonance peaks, which are hardly seen in the emission spectra for $P_r \leq 1.10$ (Fig. 4), grow rapidly 
as $P_r$ increases further above 1.10 (Fig. \ref{fig:fig7}). 
They become narrower and well separated, surpassing the stochastic emission spikes.  
There are clearly more peaks in the emission spectra with noise than those without noise at the same 
pumping level. 
This is because all modes are constantly excited by the noise and subsequently amplified in the 
presence of population inversion. 
Hence, the pump energy is distributed over more peaks. 
Nevertheless, all the lasing peaks without noise correspond to strong emission peaks with noise.  
We enumerate six major peaks in Fig. \ref{fig:fig7}(d), with 1 being the strongest.
For $P_r \geq 2.00$ [Figs. \ref{fig:fig7}(f) and \ref{fig:fig7}(g)], 
the difference between the emission spectra with noise and those without noise is reduced. 

\begin{figure}
\begin{center}
  \begin{tabular}{m{.5cm}m{7cm}}
    (a)\vspace{4cm} & \includegraphics[width=7cm]{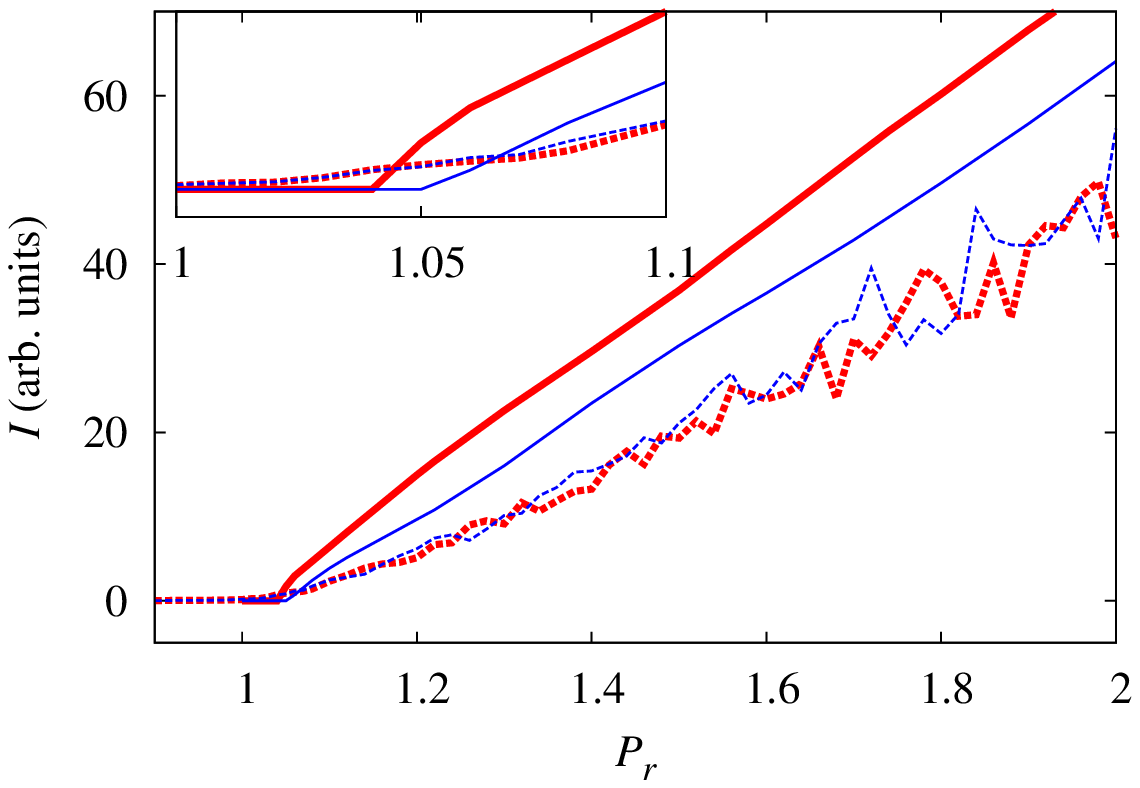}\\
    (b)\vspace{4cm} & \includegraphics[width=7cm]{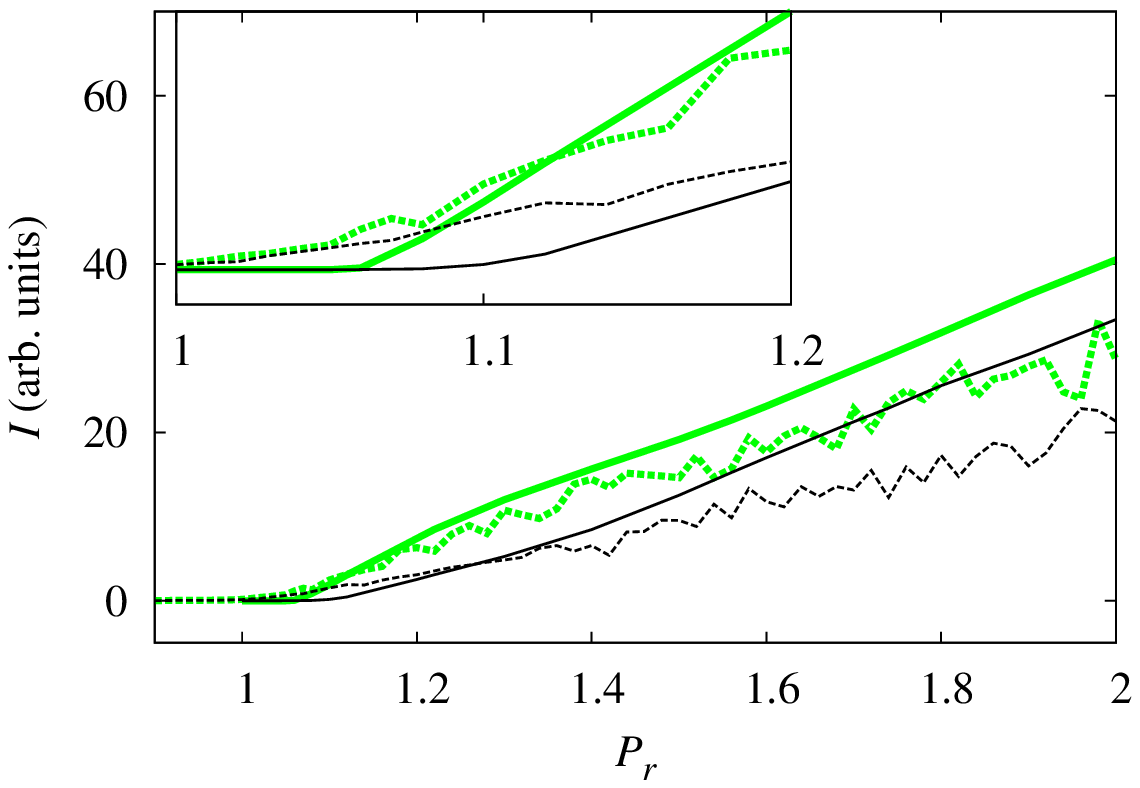}\\
  \end{tabular}
  \caption{\label{fig:fig8} (Color online)
    Modal intensities with noise (solid lines) compared to those without noise (dashed lines) for 
    a random system with $g=1.0$.
    (a) Intensity of mode 1 $I_1$ (red thick lines) and mode 2 $I_2$ (blue thin lines) vs. $P_r$.
    (b) Intensity of mode 3 $I_3$ with (green thick lines)
    and mode 4 $I_4$ (black thin lines) vs. $P_r$.
    Modes are enumerated in Fig. \ref{fig:fig7}(d). 
  }
\end{center}
\end{figure}

We compare the intensities and frequencies of the lasing modes with noise to those without noise. 
In both cases, the strongest peaks are 1 and 2 (Fig. \ref{fig:fig7}). 
We plot their intensities versus $P_r$ in Fig. \ref{fig:fig8}(a). 
In the absence of noise, there is a clear threshold for lasing. 
For example, mode 1 has zero intensity for $P_r < 1.06$. 
Once $P_r$ exceeds 1.06, its intensity rises quickly. 
The sharp turn-on at $P_r = 1.06$ marks the lasing threshold for mode 1. 
Mode 2 reaches its lasing threshold by $P_r=1.08$ and its intensity increases almost linearly with $P_r$.
Although modes 1 and 2 display notable frequency pulling just above the lasing threshold, their 
frequencies do not shift significantly as $P_r$ increases further above the threshold. 
This is due to gain saturation. 
At $P_r > 1.10$, the center frequencies of lasing peaks with noise are almost identical to those 
without noise. 
Thus noise does not affect the frequencies of lasing modes. 
Since each lasing peak has a finite width, we integrate the emission intensity over a spectral range 
set by the mid frequencies between adjacent peaks. 
As shown in Fig. \ref{fig:fig8}(a), the intensities of modes 1 and 2 increase gradually with $P_r$. 
The soft turn-on makes it difficult to pinpoint the exact value of the lasing thresholds. 
Because of ASE, the modal intensity is non-zero below the lasing threshold. 
Above the threshold pumping rate for lasing without noise, the intensity with noise is notably lower 
than that without noise because some pump energy is diverted to other modes via ASE. 
The superlinear increase of modal intensity around the threshold is caused by ASE. 
It is evident that noise almost equalizes the intensities of modes 1 and 2, despite the fact that mode 1
is clearly stronger than mode 2 without noise.       

Figure \ref{fig:fig8}(b) shows the intensity of modes 3 and 4 [enumerated in 
Fig. \ref{fig:fig7}(d)] with and without noise as $P_r$ increases.
Without noise, mode 3 reaches its lasing threshold at $P_r=1.08$.
Mode 4 has a similar lasing threshold, but its amplitude remains small until $P_r=1.22$ 
[Fig. \ref{fig:fig7}(b)].
With noise, the intensities of both modes start rising from zero at $P_r < 1.08$. 
They increase superlinearly with $P_r$ and are greater than the intensities without noise even about the
threshold for a small range of pumping rates [inset of Fig. \ref{fig:fig7}(b)].  
This is most noticeable for mode 4 in the range $P_r < 1.26$. 
The co-existence of multiple modes and their interactions through the gain material make it difficult to
define the lasing threshold for each mode using previously developed methods for single mode lasers 
\cite{jin94,ricec,straufPRL06,ulrichPRL07,beveratos}. 
Though the lasing threshold is not precisely defined here,
the soft turn-on and subsequently smaller intensities at larger $P_r$ in the case with noise shows
the threshold is increased for each of the four dominant modes (1-4) when noise is included.

\begin{figure}
  \includegraphics[width=7cm]{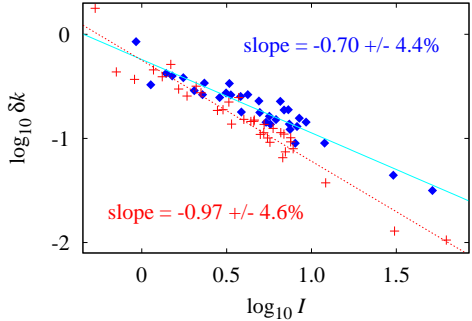}
  \caption{\label{fig:fig9} (Color online)
    Linewidths of lasing mode 1 $\delta k_1$ (red crosses) and  
    lasing mode 2 $\delta k_2$ (blue diamonds) versus 
    the corresponding mode intensity $I_1$ and $I_2$ 
    for a random system with $g=1.0$. 
    Linear fits to the data give the power by which the linewidths decrease.
    Mode 1 $\propto I_1^{-0.97}$. 
    Mode 2 $\propto I_2^{-0.70}$.
  }
\end{figure}

Next we calculate the spectral width of lasing modes, that is impossible to do with the noiseless simulation.
Considering the noisiness of the spectrum, we again use a Lorentz error function [Eq. (\ref{eq:lef})]
to obtain the linewidth objectively.
The integration of emission intensity for the Lorentz error function is limited to the spectral range of
each mode, which is the same as that used to obtain the spectrally-integrated intensity.
Figure \ref{fig:fig9} plots the linewidths $\delta k$ of modes 1 and 2 with respect to the steady-state
intensities $I$.
Mode 1 narrows the most dramatically; its linewidth decays over two orders of magnitude.
On a log-log scale, the data for each mode falls onto a straight line, indicating a power-law decay. 
We fit the data by $\delta k \propto I^{\alpha}$ within a range $I_l < I < I_u$. 
$I_l$ is set by the threshold pumping rate without noise, at which separate resonance peaks 
emerge in the presence of noise.
$I_u$ is determined by the pumping rate at which an accurate estimate of the linewidth is no longer 
possible due to limited spectral resolution (determined by the running time of the simulation).
For mode 1, the exponent $\alpha = -0.97 \pm 4.6 \%$ is close to the Schawlow-Townes prediction of 
laser linewidth \cite{st58}. 
For mode 2, $\alpha = -0.70 \pm 4.4 \%$ so the linewidth decays slower, 
probably due to mode competition for gain. 
It is known that multimode operation affects laser linewidths \cite{conti08,conti10}.
A more quantitative investigation will be carried out in the future.

Finally we look at some of the smaller peaks, e.g., 5 and 6 in Fig. \ref{fig:fig7}(d). 
Both have corresponding peaks in the noiseless spectra, but they are orders of magnitude smaller than 
the main peaks and cannot be seen on the vertical scale of Fig. \ref{fig:fig7}.
With noise, peak 6 has a much larger amplitude and is visible together with the major peaks in the 
emission spectrum.  
Close by the frequency of peak 5, there are two resonances, one on either side of $k_a$ in 
Fig. \ref{fig:fig2}(b).
At lower pumping rates, only a ``composite'' peak appears at $k= 10.3$ $\mu$m$^{-1}$. 
The linewidths of the two modes exceed their frequency spacing, which is reduced by the frequency 
pulling effect. 
Consequently, the two modes are indistinguishable and appear to be merged.   
At higher pumping rates, their linewidths decrease further, but the amplitudes remain relatively small 
compared to the four main peaks. 
Such weaker modes are affected more by gain nonlinearity, and display complicated behavior with 
increasing pumping rate. 
Detailed investigation of this behavior will be left for future studies.

\section{ASE and Lasing in a Random System with Non-overlapping Resonances\label{sec:nonovlp}}

\begin{figure}
\begin{center}
  \includegraphics[width=4.25cm]{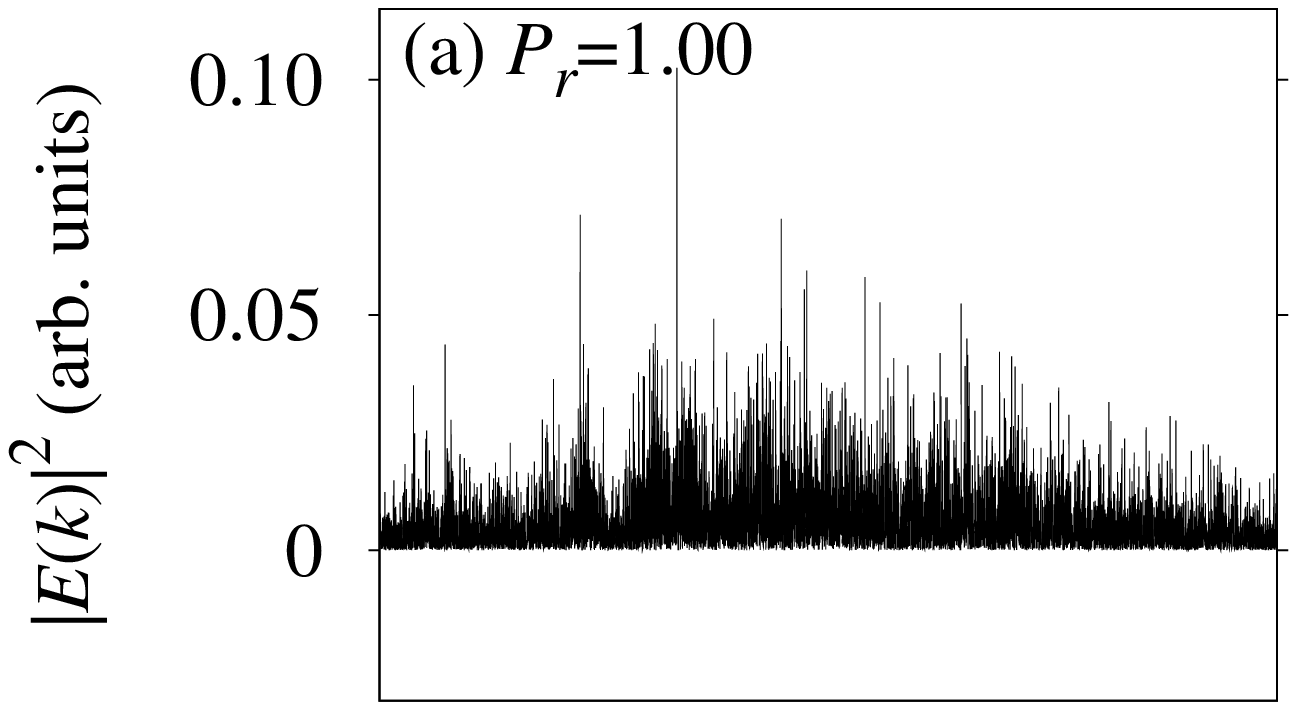}
  \includegraphics[width=4.25cm]{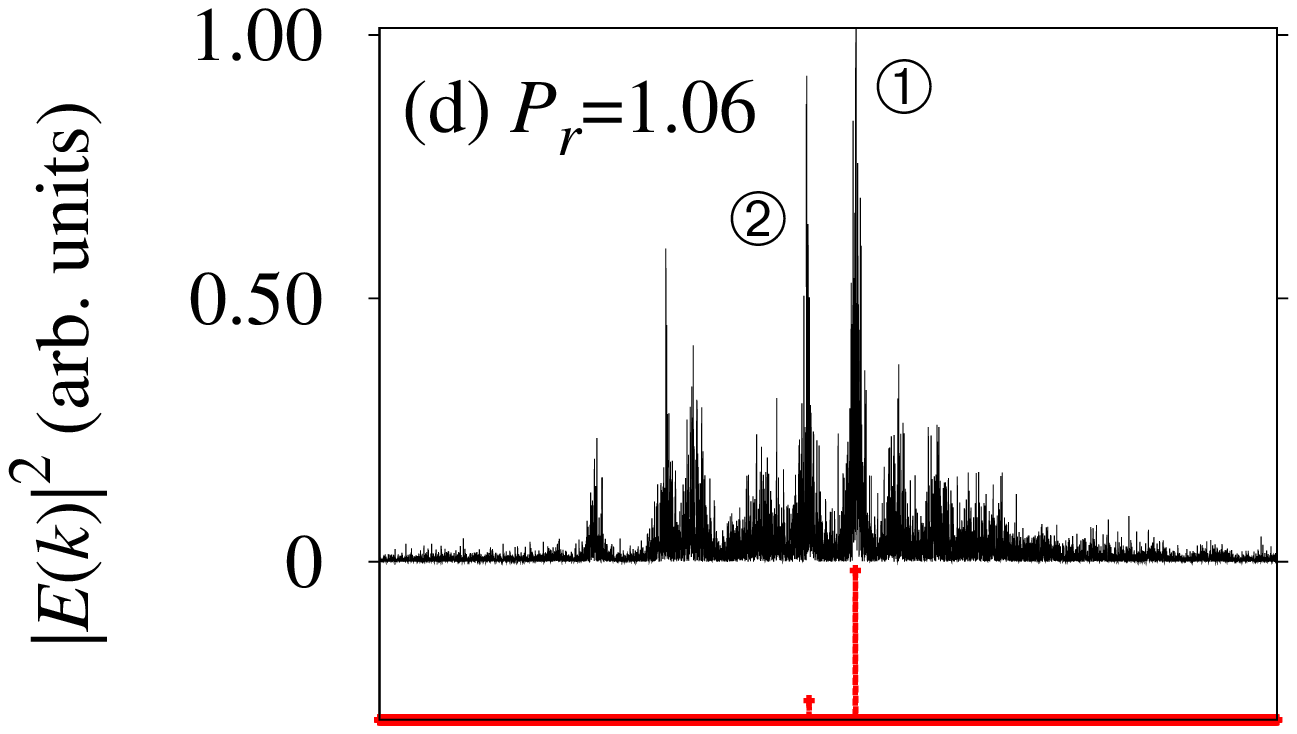}\\
  \includegraphics[width=4.25cm]{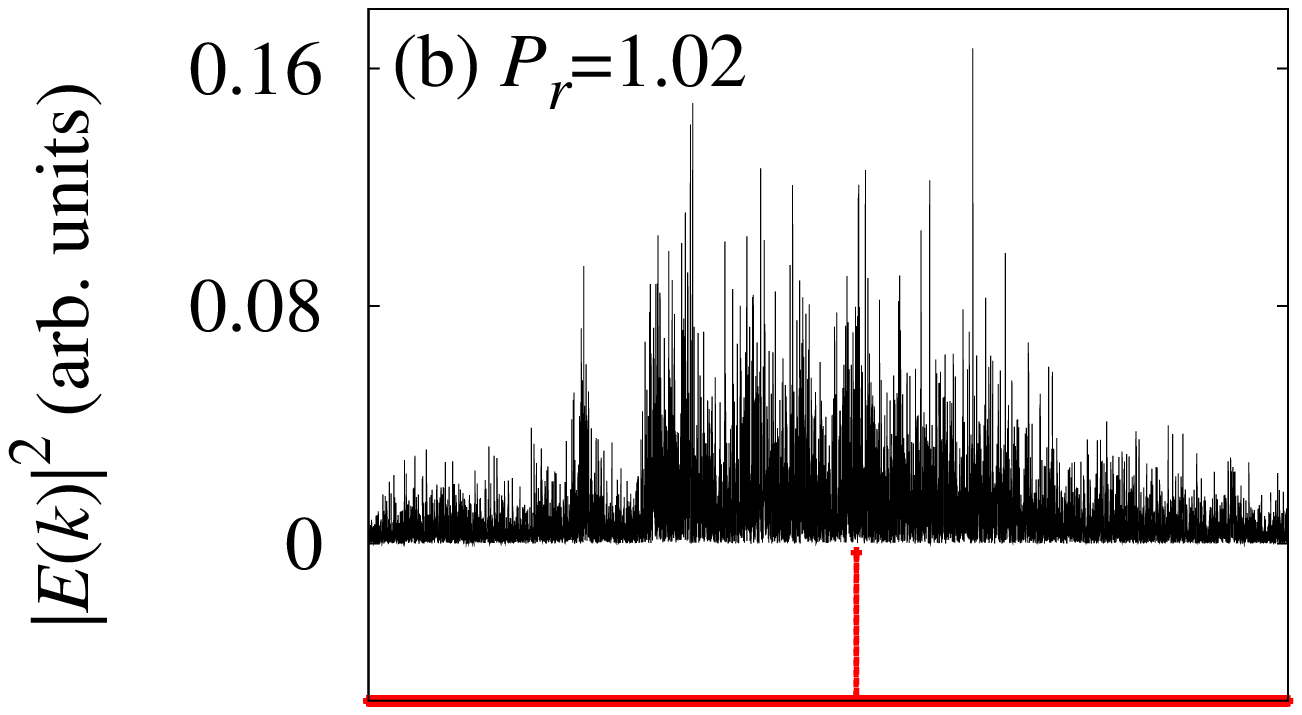}
  \includegraphics[width=4.25cm]{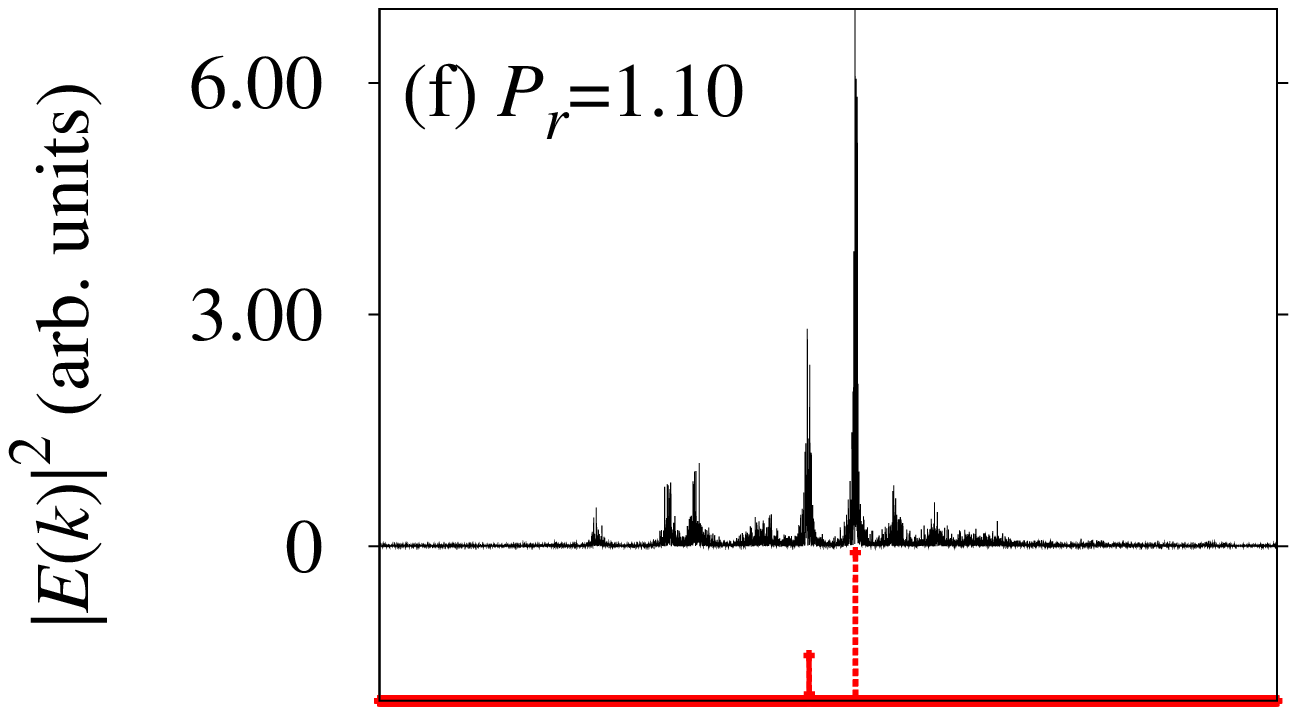}\\
  \includegraphics[width=4.25cm]{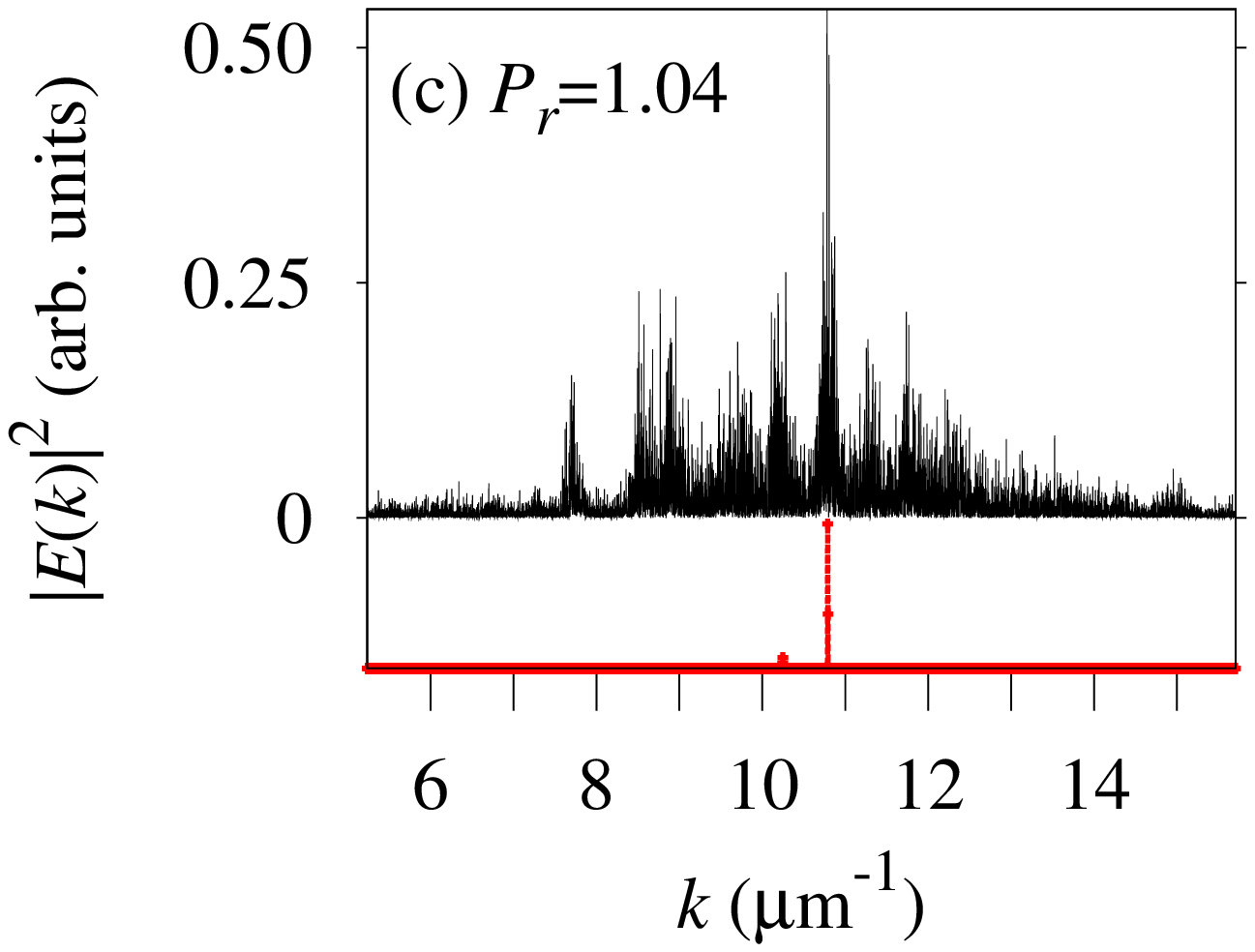}
  \includegraphics[width=4.25cm]{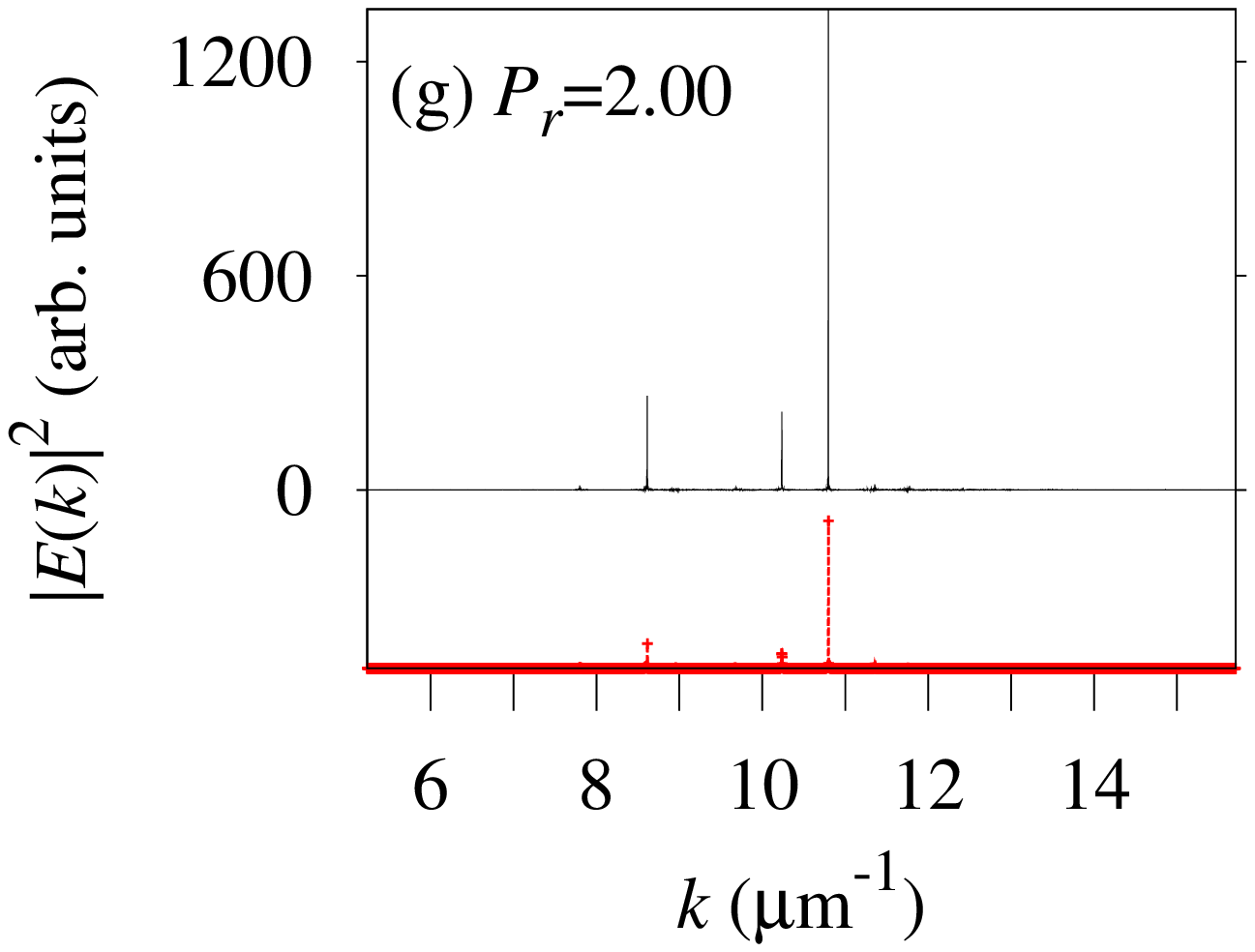}\\
  \caption{\label{fig:fig10} (Color online)
    Steady-state emission spectra $|E(k)|^2$ with noise (upper black line in each panel) compared to those 
    without noise (lower red line and crosses in the same panel) at the same pumping rate $P_r$ for a random 	    
    system with $g=0.5$. The values of $P_r$ are written in each panel. 
  }
\end{center}
\end{figure}

In this section we study laser emission characteristics of the 1D random system with $g=0.5$. 
With higher refractive index contrast ($\Delta n = 0.25$), light leakage is reduced and so is the lasing threshold. 
Figure \ref{fig:fig10} shows the steady-state emission spectra for increasing pumping 
rates with and without noise. 
$P_r$ is normalized to the value at which $\rho_3 = 0$ in the absence of noise. 
In Fig. \ref{fig:fig10}(a), $P_r = 1$ and there is no gain, so without noise the steady-state emission 
intensity is zero.  
With noise, the steady-state emission spectrum has a broad peak around the atomic transition frequency. 
Spectral modulation of emission intensity is evident. 
Higher emission intensities match the transmission peaks in Fig. \ref{fig:fig1}; 
lower intensities match the transmission dips. 
Because the Thouless number is less than unity, the quasimodes are already separated.
Particular modes may be even narrower and farther apart. 
They appear as peaks in the emission spectrum even without gain ($P_r = 1.0$).

With only a slight increase of the pumping rate to $P_r=1.02$ [Fig. \ref{fig:fig10}(b)],
a single lasing peak appears at $k=10.8$ $\mu$m$^{-1}$ in the absence of noise.  
With noise present, the broad emission peak grows and narrows around $k_a$. 
Intensity modulation is enhanced, as the resonance peaks become narrower by light amplification.  
For $P_r\ge 1.04$ [Figs. \ref{fig:fig10}(c)--\ref{fig:fig10}(g)], well separated peaks develop in the 
emission spectra with noise. 
The two strongest emission peaks, enumerated in Fig. \ref{fig:fig10}(d), have the same frequencies as 
the lasing peaks without noise.
They correspond well to the two resonances nearest $k_a$ in the passive system 
(boxed in Fig. \ref{fig:fig2}). 
By $P_r=2.00$ [Fig. \ref{fig:fig10}(g)] the spectrum with noise resembles that without noise. 
In both cases, the number of major peaks is three. 
The influence of noise is reduced for the system of smaller $g$, because
the lower lasing threshold narrows the range of pumping rates where ASE dominates. 
With increasing $P_r$, gain saturation quickly sets in to suppress the fluctuations. 

\begin{figure}
\begin{center}
  \includegraphics[width=7cm]{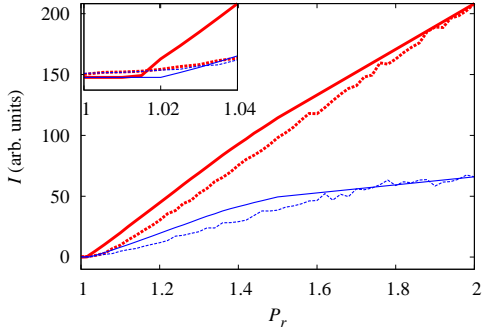}
  \caption{\label{fig:fig11} (Color online)
    Modal intensities with noise (solid lines) compared to those without noise (dashed lines).
    Intensity of mode 1 (red thick lines) 
    and mode 2 (blue thin lines)
    vs. $P_r$ for a random system with $g=0.5$.
    Modes are enumerated in Fig. \ref{fig:fig10}(d).
  }
\end{center}
\end{figure}

Figure \ref{fig:fig11} plots the spectrally-integrated intensity of the two peaks enumerated in 
Fig. \ref{fig:fig10}(d). 
Compared to the two modes in Fig. \ref{fig:fig8}(a), the increase of intensity with pumping is more rapid.
As before, we cannot pinpoint the exact lasing threshold for each mode because of multimode operation. 
Nevertheless, it is evident that without noise the onset of lasing oscillation occurs at a lower pumping 
rate, and the modal intensity is higher than that with noise. 
This is because the pump energy is partly consumed by ASE in other modes in the presence of noise. 
However, by $P_r=2.00$ the effect of noise is diminishing, and the modal intensity with noise converges 
to that without noise.  
Hence, the transition from the amplification of spontaneous emission to the lasing oscillation happens over a relatively small range of pumping rates, 
and the effect of noise is less significant than that in the random system with larger $g$. 

\section{Conclusion \label{sec:conclusion}}

The effects of fluctuations caused by interactions of atoms with reservoirs were studied in random lasers.
A FDTD-based method for solving the stochastic Maxwell-Bloch equations was employed. 
It is particularly well-suited for studies of light-matter interaction in complex systems without prior knowledge of resonant modes.
Two random systems with different degrees of spectral overlap of resonances were investigated.
We were able to simulate amplified spontaneous emission (ASE) below the lasing threshold and capture the transition from ASE to lasing. 

In the case of overlapping resonances, the emission spectra at low pumping are broad and continuous. 
Above the transparency point, frequency-selective amplification leads to a dramatic narrowing of the emission spectrum and a superlinear 
increase of the peak emission intensity. 
Such behavior is in accordance with early experimental results \cite{lawandy94}.   
Moreover, our simulation reproduced the stochastic emission spikes in the spectra, with similar characteristics to the experimentally 
observed ASE spikes \cite{wu07OL, wu08}.  
Previous experiments found the spectral width of ASE spikes depends on the temporal duration of the emission pulse.
Here, we found the width of stochastic emission spikes is determined by the integration time of the Fourier
transform of the output field.
The statistical distribution of frequency spacing of spikes displays an exponential tail, as seen experimentally. 
The spikes have no relation to the resonant modes of the system, 
and can be clearly differentiated from the emission peaks formed by resonances.

We compared the lasing behavior with noise to that without noise in the same random system. 
The lasing peaks in the spectra without noise coincide with peaks in the spectra with noise.
Hence, noise does not affect mode frequencies.
However, all modes within the gain curve are constantly excited by noise and subsequently amplified by stimulated emission.
Therefore, there are always multiple modes appearing in the steady-state emission spectra. 
The regime of single mode lasing, realized in the noiseless simulation by fine tuning of pump, disappears. 
With some portion of pump energy diverted to ASE in other modes, the lasing modes have higher thresholds than those without noise.  
Moreover, the ASE below the lasing threshold results in a soft turn-on of the lasing mode.
It is in sharp contrast to the abrupt turn-on in the simulation without noise, where the emission intensity vanishes below the lasing threshold. 
When the pumping rate is well above the threshold value, the spectra with noise become more similar to the 
spectra without noise, both showing multimode lasing.
Thus, noise has the greatest influence on lasing behavior near threshold.
With noise included, we can calculate the spectral widths of individual lasing modes, and observe their decrease with increasing pump. 
The decrease appears to follow the Schawlow-Townes law for the strongest lasing mode, but not for other modes, probably due to mode interactions.

The effects of noise on lasing become less significant in random systems with a smaller degree of spectral overlap of resonances.
If the Thouless number is less than unity, the resonant modes can be resolved in the emission spectra below the transparency point. 
ASE narrows the resonance peaks, making them more distinct.   
The transition from ASE to lasing occurs over a narrower range of pumping rate, 
because the lasing threshold is lower in the case of non-overlapping modes.  
The difference between the simulation results with less mode overlap and those with larger mode overlap agrees qualitatively to the experimental data 
\cite{cao00} that compare different particle densities.
Increasing the refractive index contrast $\Delta n$ in our simulation enhances the scattering strength, which is similar to increasing the density of 
scattering particles in the experiments.

These studies shed light on the transition from ASE to lasing in random systems, that is poorly understood.
The results presented here are limited to the steady state. 
Noise is expected to have a greater effect on the dynamics, e.g., the buildup of lasing modes and
temporal fluctuations and switching of lasing modes on short time scales.
These phenomena can be studied with our numerical method.  
Furthermore, this method can be extended to the study of random lasing in higher dimensions.  
The larger density of modes and potentially stronger mode overlap in frequency may enhance the noise effects.

\begin{acknowledgments}
  The authors thank Allen Taflove, Prem Kumar, Chang-qi Cao, Christian Vanneste,
  and Patrick Sebbah for stimulating discussions.
  This work was supported partly by 
  the National Science Foundation under Grants No. DMR-0814025 and DMR-0808937 
  and by 
  the facilities and staff of the Yale University 
  Faculty of Arts and Sciences High Performance Computing Center.
\end{acknowledgments}


\end{document}